\documentclass[lettersize,journal]{IEEEtran}
\usepackage{amsmath,amsfonts}
\usepackage{algorithmic}
\usepackage{algorithm}
\usepackage{array}
\usepackage{tabularx}
\usepackage[caption=false,font=normalsize,labelfont=sf,textfont=sf]{subfig}
\usepackage{textcomp}
\usepackage{stfloats}
\usepackage{url}
\usepackage{verbatim}
\usepackage{graphicx}
\usepackage{cite}
\usepackage{xcolor}
\usepackage{booktabs}
\usepackage{orcidlink}

\usepackage{array}
\usepackage{multirow}
\usepackage{makecell} 
\begin{document}

\title{3D Stack In-Sensor-Computing (3DS-ISC): Accelerating Time-Surface Construction for Neuromorphic Event Cameras}

\author{IEEE Publication Technology,~\IEEEmembership{Staff,~IEEE,}

\thanks{This paper was produced by the IEEE Publication Technology Group. They are in Piscataway, NJ.}
\thanks{Manuscript received April 19, 2021; revised August 16, 2021.}}

\author{
    \IEEEauthorblockN{
        Hongyang Shang\orcidlink{0009-0007-6276-1947},
        Shuai Dong\orcidlink{0009-0007-4807-5094},
        Ye Ke\orcidlink{0009-0002-9809-1192},
        Arindam Basu\orcidlink{0000-0003-1035-8770},~\IEEEmembership{Senior Member,~IEEE}
    }\\
    \thanks{
        Hongyang Shang, Shuai Dong and Arindam Basu are with Department of Electrical Engineering,
         City University of Hong Kong, Hong Kong (Corresponding authors: Arindam Basu, arinbasu@cityu.edu.hk).
        
    This work was sponsored in part by RGC (C7003-24Y) and Innovation technology Fund Mid-Stream Research program under Grant ITS/018/22MS.
}
}

\markboth{Journal of \LaTeX\ Class Files,~Vol.~14, No.~8, August~2021}%
{Shell \MakeLowercase{\textit{et al.}}: A Sample Article Using IEEEtran.cls for IEEE Journals}

\IEEEpubid{0000--0000/00\$00.00~\copyright~2021 IEEE}

\maketitle

\begin{abstract}
This work proposes a 3D Stack In-Sensor-Computing (3DS-ISC) architecture for efficient event-based vision processing. A real-time normalization method using an exponential decay function is introduced to construct the time-surface, reducing hardware usage while preserving temporal information. 
The circuit design utilizes the leakage characterization of Dynamic Random Access Memory(DRAM)  for timestamp normalization. Custom interdigitated metal-oxide-metal capacitor (MOMCAP) is used to store the charge and low leakage switch (LL switch) is used to extend the effective charge storage time. The 3DS-ISC architecture integrates sensing, memory, and computation to overcome the memory wall problem, reducing power, latency, and reducing area by 69$\times$, 2.2$\times$ and 1.9$\times$, respectively, compared with its 2D counterpart. Moreover, compared to works using a $16$-bit SRAM to store timestamps, the ISC analog array can reduce power consumption by three orders of magnitude. In real computer vision (CV) tasks, we applied the spatial-temporal correlation filter (STCF) for denoise, and 3D-ISC achieved almost equivalent accuracy compared to the digital implementation using high precision timestamps. As for the image classification, time-surface constructed by 3D-ISC is used as the input of GoogleNet, achieving 99\% on N-MNIST, 85\% on N-Caltech101, 78\% on CIFAR10-DVS, and 97\% on DVS128 Gesture, comparable with state-of-the-art results on each dataset. Additionally, the 3D-ISC method is also applied to image reconstruction using the DAVIS240C dataset, achieving the highest average SSIM (0.62) among three methods. This work establishes a foundation for real-time, resource-efficient event-based processing and points to future integration of advanced computational circuits for broader applications.
\end{abstract}

\begin{IEEEkeywords}
Dynamic vision sensor, Neuromorphic, 3D integration, eDRAM, Event based sensor  
\end{IEEEkeywords}

\section{Introduction} 
Dynamic Vision Sensors (DVS) or Event-based Cameras (EBC) are bio-inspired or neuromorphic imagers where each pixel operates asynchronously by triggering an event when its temporal contrast exceeds a threshold\cite{4444573}. The event is represented by the position of the pixel (x, y), the time stamp of the event (t) and the polarity (p) indicating whether the event was triggered by an increase or decrease in brightness. Compared to conventional CMOS image sensors (CIS), this new modality inherently provides data compression capabilities, enabling high-speed, low-latency data capture with low power consumption \cite{brandli2014240}. For example, EBCs can operate at effective frame rates $>1$ kHz \cite{9389490} while conventional video cameras are typically limited to $100$ Hz. This has led to widespread interest in its usage in the fields of autonomous driving, robotics, and unmanned aerial vehicle, areas where the clarity of imaging fast-moving objects is crucial\cite{gallego2020event,gehrig2024low,yang2024vision,shariff2024event}.  

Traditionally, computer vision algorithms have been developed to process videos by considering them as a series of frames \cite{gonzalez2009digital, szeliski2022computer,dong2023performance}. The current explosive growth in vision applications such as face recognition \cite{9956236}, object tracking \cite{mitrokhin2018event}, human pose detection \cite{toshev2014deeppose}, etc. spurred by deep neural networks (DNN) \cite{bengio2017deep} also use image frames as input. However, EBCs provide data as asynchronous events in Address Event Representation (AER) format; hence, traditional frame-based computer vision algorithms are not suitable to process these events\cite{ouyang2024scalable}. 
Hence, specialized algorithms have been developed to tackle this challenge \cite{wang2019space, almatrafi2020distance,bi2019graph,6589170, 6112233}. Early methods stored the events in a queue according to time of arrival leading to a memory-efficient data structure. However, this needed a long search time due to lack of spatial information. Consequently, new 2D representations such as Surface of Active Events (SAE)\cite{benosman2013event} or Time-Surfaces (TS) \cite{lagorce2016hots} were developed to capture both spatial and temporal information. An added advantage of this representation is that it can be processed directly by existing computer vision algorithms including deep neural networks to accomplish more complex tasks, while retaining the enhanced temporal precision offered by EBCs. 
\IEEEpubidadjcol

The  SAE is a 2D data structure that retains the time of the most recent event for each pixel location and for each polarity \cite{benosman2013event}. However, SAE is theoretically unbounded due to increase in timestamp values and suffers from reset issues when the counter in practical systems wraps around. Based on SAE, \cite{lagorce2016hots} described a normalized representation called TS that uses an exponentially decaying kernel to provide temporal memory at each pixel location. The TS is a very popular representation and has been used for stereo vision \cite{zhou2021event}, object recognition \cite{kim2021n}, face recognition \cite{ryan2023real, lagorce2016hots} and other applications.
However, efficient hardware implementations of TS are lacking. Earlier works\cite{lagorce2016hots} created the TS on a computer requiring all events to be sent off-chip. This results in significant energy and latency penalties. Navarro et al. reported a system where on-chip SRAM was utilized to maintain a Timestamp+Polarity Image (TPI), recording the latest timestamps (in milliseconds) and the corresponding ON or OFF brightness change events for each DVS pixel \cite{rios2023within}. While it reduces communication energy with an off-chip memory, the power and area usage of multi-bit SRAM remains substantial, and the issue of timestamp overflow was not considered. Another work uses the simulated characteristics of an Electrochemical RAM (ECRAM) device to mimic the exponential decay required in TS, but suffers from large write energy of the memristor \cite{rasetto2023building}. Also, these devices are not yet widely available in CMOS compatible high-yield forms.

In this work, we propose an EBC processor based on embedded Dynamic Random Access Memory (eDRAM) and 3D stacking technology to enable an in-sensor computing (ISC) architecture for building TS and apply the TS on denoise, classification and image reconstruction tasks. The main contributions of this work are as follows:
(1) We propose an in-sensor computation architecture applied to 3D stacking technology (3DS-ISC), greatly improving the chip power efficiency, throughput and reducing latency by minimizing data movement.
(2) SPICE simulations are conducted using TSMC's 65nm technology to analyze the circuits' performance and non-ideal characteristics.
(3) The results were compared with 2D architecture, showing 69x lower power, 2.2$\times$ lower latency, and 2$\times$ higher area efficiency. The results also demonstrated that the ISC analog array reduces power consumption by 1600-6761$\times$ and area overhead by 2.2-3.1$\times$, compared to $16$-bit SRAM-based timestamp storage implementations.
(4) Validation is conducted using commonly used datasets, including DND21 for denoising, N-MNIST, N-Caltech101, CIFAR10-DVS, and DVS128 Gesture for classification, and the DAVIS240C dataset for image reconstruction.

The remainder of the paper is organized as follows. Section II introduces relevant concepts and reviews prior work in the domain. Section III describes the co-design of the exponential decay TS and its hardware implementation. Section IV presents the hardware performance, compares it with traditional architectures, and demonstrates its application in denoising, classification and image reconstruction tasks. Finally, Section V concludes the paper.
 
\section{PRELIMINARIES AND RELATED WORKS}
\label{sec:prelim}
\subsection{3D integration for In-sensor Computing}
With the rise of in-sensor and near-sensor computations, challenges related to data movement-induced latency and energy loss have been addressed. In traditional von Neumann architecture, input data is stored in memory, such as SRAM, waiting for control instructions before being sent to the CPU for processing. However, in-sensor computing architectures can perform calculations simultaneously while accessing memory, thus solving the memory bottleneck problem \cite{bose2022389}. The advent of 3D stacking technology using through-silicon vias (TSV) and microbumps further promotes the development of in-sensor computation by putting the memory closer to the sensor. Specifically, only $\approx 0.7$ fJ of energy is consumed to transmit one byte of data by Cu-Cu bonding \cite{ku2018area}, much less than the energy needed to transmit video data over traditional interconnects using traditional interconnects, such as peripheral component interconnect express (PCIe) and MIPI CSI-2. 3D stacking has been used to increase the fill factor in EBCs \cite{guo2023three} by placing the event detection circuits in a separate wafer below the one containing the photodiodes. However, the usage of 3D stacking to closely integrate processing with EBCs has not yet been explored.

\begin{figure}[t]
    \centering
    \includegraphics[width=0.45\textwidth]{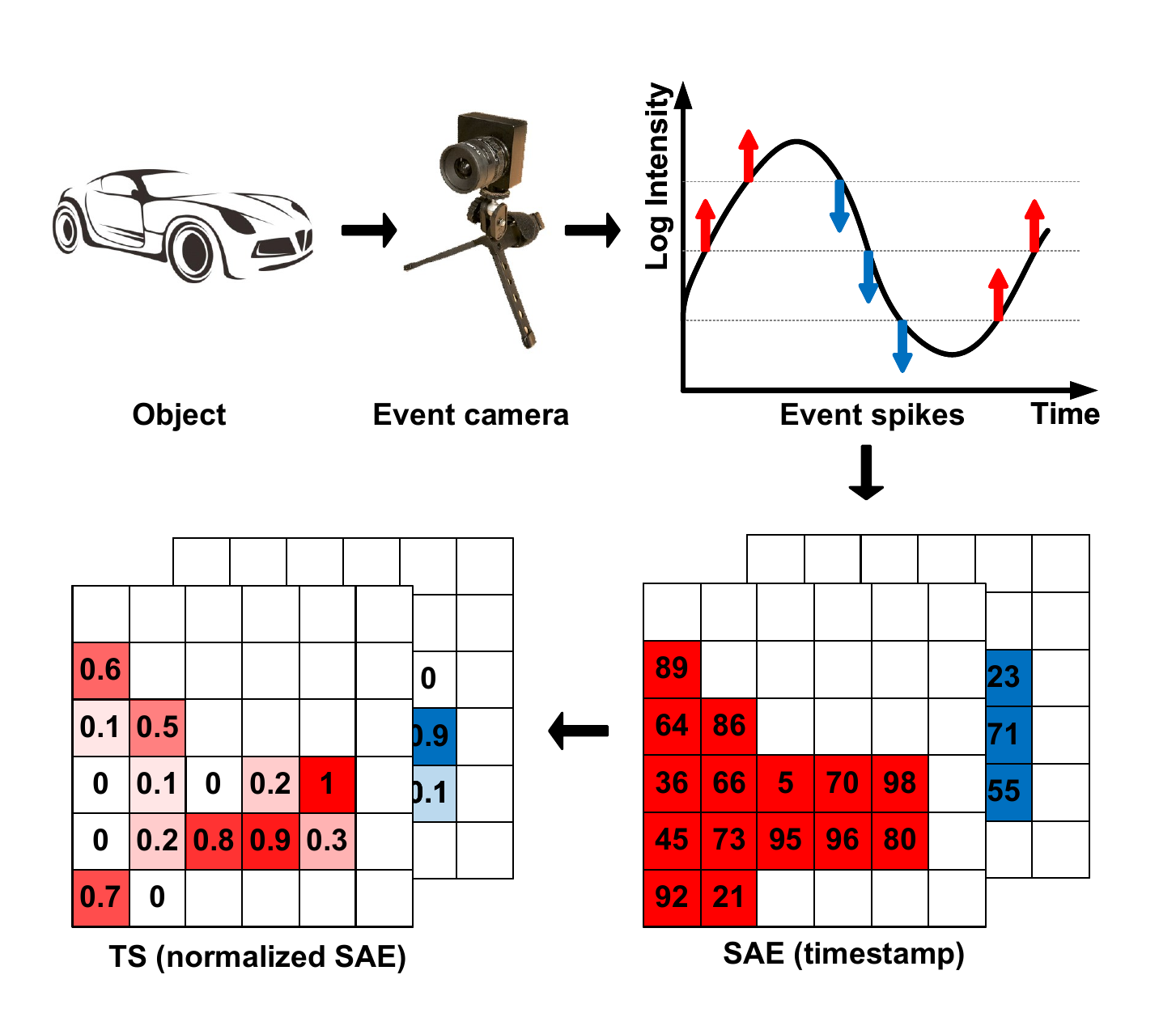}
    \caption{
    Concept of event camera's time-surface (TS). DVS generates event pulses based on changes in temporal contrast. Timestamps of each event are subjected to exponential decay using equation (\ref{eq:TS}), creating a 2D representation termed as the TS.}
    \label{fig:concept_ts} 
\end{figure}

\subsection{2D Event Data Representations}
As described earlier, DVS/EBC is a novel type of vision sensor that generate streams of events by detecting changes in pixel-wise temporal contrast\cite{posch2010qvga,4444573} (though some imagers may also sense changes in absolute intensity) in the scene (see Fig. \ref{fig:concept_ts}). Each event is represented by its pixel location, timestamp, and polarity of the intensity change. The event stream generated by DVS can be represented by a mathematical expression: 

\begin{align}
\label{eq:eq1}
    \mathrm{E}&=\{e_i\}_I \notag\\
    e_i&=\left[x_i, y_i, t_i, p_i\right], i \in \mathrm{I}
\end{align}
where $x_{i}$ and $y_{i}$ represent the coordinate of the $i^{th}$ event ($e_{i}$), $t_{i}$ is the timestamp, and $p_{i}$ is the polarity. Different data representations have been used to store event data. They vary in terms of their information content and resource (memory/power) usage. The simplest representations are event count based images, where the number of events occurring at a pixel is summed over a frame duration \cite{liu2018adaptive,maqueda2018event}. The event count representation demands moderate memory, quantified as H × W ×  $n_C$ bits, where H and W correspond to the pixel array's row and column dimensions, respectively, and $n_C$ represents the bit width of the counter (typically $n_C$ $\approx$ 4). The other problem with this approach is it loses the high temporal precision of event cameras necessary for high-speed vision. A simplified version of this approach is event-based binary image (EBBI) where this count is binarized to reduce memory usage \cite{acharya2019ebbiot,mohan2022ebbinnot}. This method achieves a memory complexity of O(H × W). However, it still suffers from the same problem of reduced temporal precision.

Alternatively, a temporal representation termed as surface of active events (SAE) was proposed in \cite{mueggler2017fast}. SAE stores the timestamp of the most recent event at each pixel location as shown in Equation \ref{eq:SAE}: 
\begin{equation}
\label{eq:SAE}
    \operatorname{SAE}\left(x_i, y_i, p_i\right) = t_i
\end{equation}
This is the most basic method for converting event data into 2D format, while retaining the temporal coherence of incoming events. However, the high memory requirement of H × W × $n_T$ bits results in increased energy consumption and greater area overhead, where $n_T$ represents the bit-precision of the timestamp stored in memory (typically $n_T$ $\geq$ 16). SAE has also been used to create TS \cite{lagorce2016hots}. To create a TS, the pixel timestamp values in a patch surrounding the active event ($x_k$,$y_k$,$t_k$) are converted to analog values following an exponential decay given by: 
\begin{equation}
\label{eq:TS}
TS_k(u, p) = e^{-\frac{t_k - T_k(u, p)}{\tau}}
\end{equation}
Here, u is the surrounding patch of the incoming event $e_k$, and $T_k$ (u,p) is a timestamp patch provides the time-context around $e_k$. While SAE is unbounded, the TS is naturally normalized to a maximum value of 1. An improved version of TS, termed Local Memory Time-Surface, was introduced in \cite{sironi2018hats} that reduced the sensitivity of TS to noise events. This was done by creating a weighted summation of decaying exponentials centered at every event in a past time window.  
The SAE/TS representation has been used for image recognition \cite{lagorce2016hots}, face recognition \cite{lagorce2016hots,ryan2023real}, optical flow \cite{nagata2021optical}, image reconstruction \cite{scheerlinck2020fast}, visual odometry \cite{zhu2023event}, etc. 

More recently, two other representations have been introduced – Speed Invariant Time Surface (SITS) \cite{manderscheid2019speed} and Time Ordinal Surface (TOS) \cite{glover2021luvharris}.
But the problem with both SITS and TOS is that they require far more memory writes ($ \sim $25-50$\times$ number of events) compared to the previous approaches where the number of memory writes was equal to the number of events. This makes it difficult to implement these approaches in low-energy and low-latency hardware.

Next, we describe hardware implementations of TS/SAE and their corresponding challenges.

\subsection{Hardware Architectures for Time-Surface Computation}
\subsubsection{Electro-chemical RAM (ECRAM)-Based Time-Surface Implementation}
ECRAM leverages the intrinsic volatility of memristive devices to emulate TS. A lithium-tungsten oxide (LixWO3) memristor exhibits dual-exponential conductance decay, mimicking biological short-term synaptic plasticity. Each spatial location and polarity (x, y, p) is assigned a memristor, whose conductance is set to a high value by event-triggered pulses. Conductance decay naturally generates temporal memory, replacing exponential decay calculations. By sampling memristor states during events, the ECRAM enables analog computation of TS. However, this work only uses simulation models since such diffusive memristors are not yet commercially available. Moreover, the write energy for this device is very high ($\approx 250$ nJ based on $0.5$ V read pulse and average conductance value of $100$ $\mu S$). Lastly, the write pulse duration of $10$ ms is also an impediment when working with high-frequency event pulses generated by scenes with high contrast.

\subsubsection{Timestamp-Polarity Integration in Digital Memory}

Digital implementations store the TS as a timestamp-polarity information (TPI) using SRAM \cite{rios2023within}. Each pixel’s timestamp ($16$-bits) and polarity ($2$-bits) are stored in dedicated memory banks. For a sensor with a resolution of $346\times 260$, FPGA implementations require approximately 400 BRAM blocks, while ASIC designs utilize $18$-bit SRAM banks, each containing 2,048 words, distributed across 44 banks. In the case of the ASIC design, the energy consumption for writing timestamps is $0.072$ nJ/event, with a high leakage power of $35$ mW required for storage. Additionally, the area overhead for the SRAM cells alone, using 65 nm technology, amounts to as much as $4.3$ $mm^2$ which occupies 99.5 \% of the total area. It is also important to note that the limitation related to timestamp overflow has not been addressed.

\section{Algorithm, Architecture, Circuit co-design for time-surface construction}
\label{subsec:theory}

To enable an efficient creation of a TS from the event stream of a DVS/EBC, we propose several circuit and architecture innovations as described next.

\subsection{Algorithm-Circuit Co-Design: Low-leakage 6T-1C embedded DRAM}
\label{subsec:bitcell}
The intuition for our design stems from the fact that charge-based memory cells such as DRAMs naturally encode the time elapsed since the last write event in the residual voltage on the storage capacitor. Embedded DRAMs are also known to be more area efficient than SRAMs and have been used in several recent In-memory computing efforts\cite{bongjin_edram}. Hence, it is feasible to assign a eDRAM cell to each pixel to store its TS value. If a logic high voltage denoted by $V_{reset}$ is written to the eDRAM for pixel ($x_i,y_i$) at the time of the event $t_i$, its voltage can be written as:
\begin{align}
\label{eq:dram}
    V_{(x_i,y_i)}(t)&=V_{reset}\text{ for }t=t_i\notag\\
     V_{(x_i,y_i)}(t)&=f(V_{reset},t-t_i,\tau)\text{ for }t>t_i
\end{align}
where $V_{reset}$ is nominally equal to $V_{dd}$ and $\tau$ represents the decay time constant. This TS offers two distinct advantages. First, it directly normalizes time by a storing it as a physical voltage where the most recent time is denoted by $V_{reset}$. This avoids the storage of high-precision timestamps, thus significantly conserving memory space. This also overcomes timestamp overflow issues related to finite bit-width of digital number representation. Second, this method retains more information compared to binary images since relative timing of events is preserved. Assuming an exponential form of the function $f()$, this method can create a TS given by the following equation:
\begin{equation}
\label{eq:SAE_global}
    T S\left(x_i, y_i, p_i\right)=e^{-\frac{t-\operatorname{SAE}\left(x_i, y_i, p_i\right)}{\tau}}
\end{equation}
where we use the previous definition of SAE in equation (\ref{eq:SAE}). Note that in previous digital implementation of TS, the exponential calculations are done only in a neighbouring patch of an event when necessary, since performing exponential operations on the entire image all the time is extremely costly. In contrast, this exponential decay happens naturally and parallely across the entire eDRAM aray in our proposed case. We show later that the exact function $f()$ for our eDRAM is better modelled by a double exponential function which also performs well in maintaining accuracy similar to the original TS in all the tasks. In fact, any $f()$ that can capture time elapsed since last event write should work well.

\begin{figure}[htbp]
    \centering
    \includegraphics[width=0.5\textwidth]{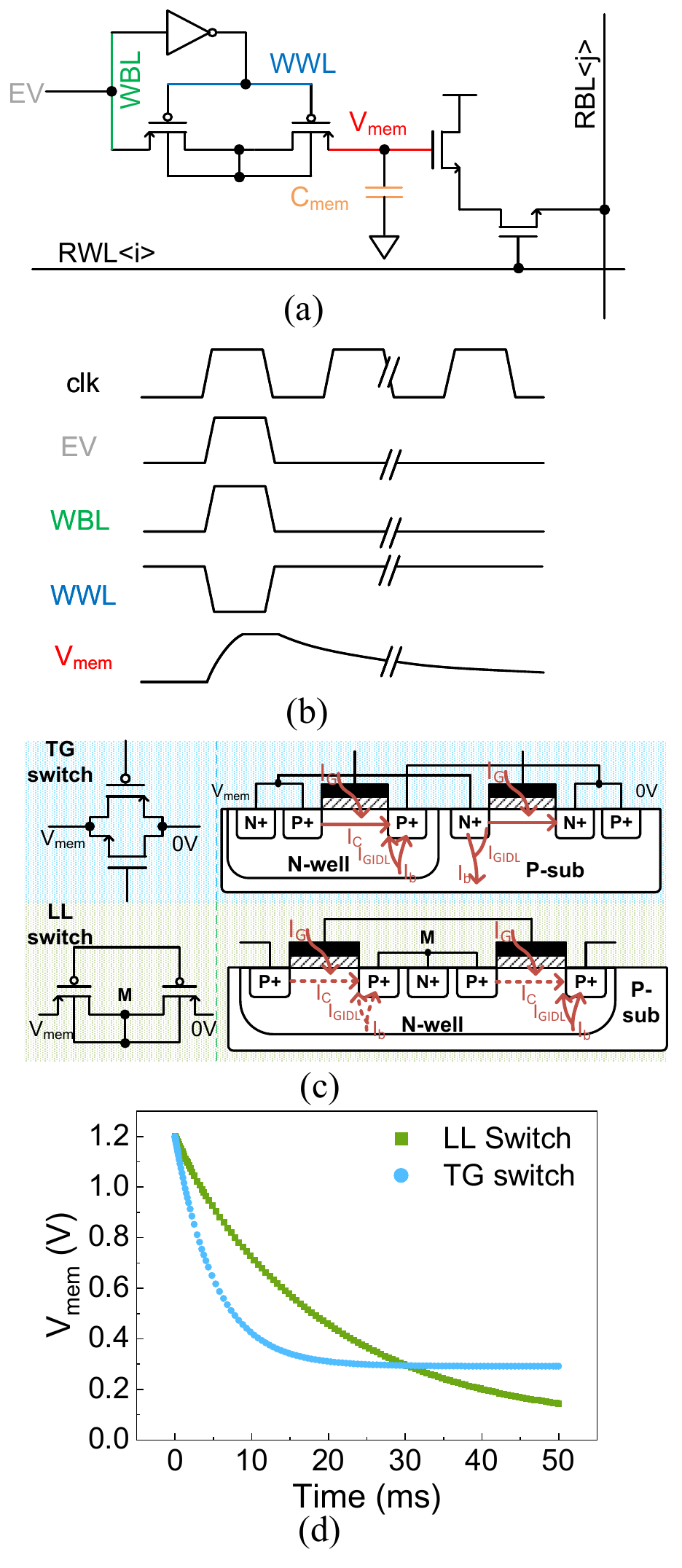}
    \caption{The 6T-1C eDRAM cell design and comparison of leakage performance between the TG and LL switch. (a) The eDRAM cell using a low-leakage switch comprising two PMOS transistors with floating well. The EV signal from the EBC is connected to the Write-bit line (WBL) of the eDRAM. (b) Since the WBL is directly connected to the EV, a positive pulse appears on the WBL while the inverter creates a negative pulse on the write word line (WWL). This completes the event writing process. (c) Circuit diagrams and cross-sectional views of the TG and LL switch. In the LL switch structure, the stack of two PMOS transistors reduces the $V_{ds}$, thereby decreasing the channel leakage current ($I_{c}$) in the off state (indicated by the dashed line). (d) The decrease in $V_{mem}$ over time using both switch types is displayed.  
    }
    \label{fig:dram_cell} %
\end{figure}

The remaining issue stems from the relatively small retention time values of $\approx 200$ $\mu$s achievable in eDRAM technology\cite{bongjin_edram}. This can be solved by using a low-leakage (LL) switch used to create large pseudo-resistors in neural recording amplifiers\cite{reid_pseudo}. However, unlike conventional pseudo-resistors, the switch in our design needs to be able to switch to a low resistance mode as well to facilitate low-latency write operations. Fig. \ref{fig:dram_cell}(a) depicts the proposed 6T-1C eDRAM cell where two PMOS transistors with floating wells are used as a LL switch. The other two transistors are used in the inverter used to drive the Write Word Line (WWL) which turns the switch ON or OFF. The last two NMOS transistors are used for selection and source-follower based readout, respectively, similar to active pixel image sensors. This is because unlike the digital data stored in other eDRAM bitcells shown in Table \ref{tab:memory_tech}, our design has analog data whose readout mechanism is similar to image sensors. Table \ref{tab:memory_tech} also shows the increased memory time in our proposed cell compared to other bitcells.

\begin{table*}[h]
\centering
\caption{Comparison of different types of DRAM}
\label{tab:memory_tech}
\begin{tabular}{|>{\centering\arraybackslash}m{1.5cm}|  
                >{\centering\arraybackslash}m{2cm}| 
                >{\centering\arraybackslash}m{2cm}| 
                >{\centering\arraybackslash}m{2.2cm}| 
                >{\centering\arraybackslash}m{2cm}| 
                >{\centering\arraybackslash}m{2.3cm}| 
                >{\centering\arraybackslash}m{2.5cm}|} 
\hline
\multirow{2}{*}{Category} & \multicolumn{6}{c|}{Memory Technology} \\ \cline{2-7}
 & 1T1C\cite{barth2008500} & 3T\cite{chun20113t} & 2T1C\cite{chun2011700mhz} & 2T\cite{chun2011667} & 2D 4T1C & 3D 6T1C \\ 
\hline Picture 
& \makecell{\includegraphics[width=1.5cm]{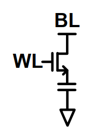}}
& \makecell{\includegraphics[width=2cm]{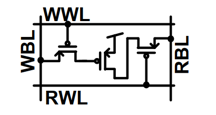}}
& \makecell{\includegraphics[width=2cm]{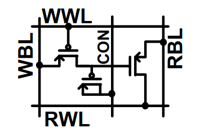}} 
& \makecell{\includegraphics[width=2cm]{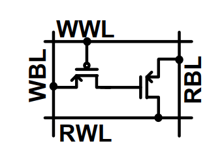}}
& \makecell{\includegraphics[width=2cm]{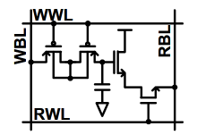}}
& \makecell{\includegraphics[width=2cm]{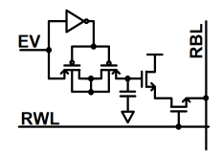}} \\

\hline Data type & Digital & Digital & Digital & Digital & Analog & Analog \\ 

\hline Pros 
& \makecell{High density,\\ low power\\ consumption} 
& \makecell{Decoupled\\ read/write,\\ Standard process} 
& \makecell{Decoupled\\ read/write,\\ Standard process,\\no boosted \\supplies} 
& \makecell{Decoupled\\ read/write,\\ Standard process} 
& \makecell{Decoupled\\ read/write,\\ Standard process,\\ high retention\\ time} 
& \makecell{Decoupled\\ read/write,\\ Standard process,\\ high retention\\ time} \\ 
\hline Cons
& \makecell{Destructive read,\\ Half selection\\ issue, require\\ deep trench\\ capacitor} 
& \makecell{Half selection\\ issue, need\\ boosted supplies,\\ low retention\\ time} 
& \makecell{Half selection\\ issue, \\ low retention\\ time} 
& \makecell{Half selection\\ issue, need\\ boosted supplies,\\ low retention\\ time} 
& \makecell{Half selection\\ issue} 
& \makecell{Relative low\\ density} \\ 
\hline Leakage 
& \makecell{\includegraphics[width=2cm,height=2cm]{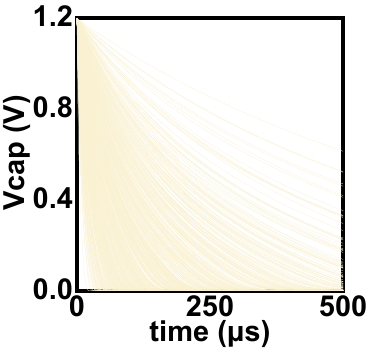}}
& \makecell{\includegraphics[width=2cm,height=2cm]{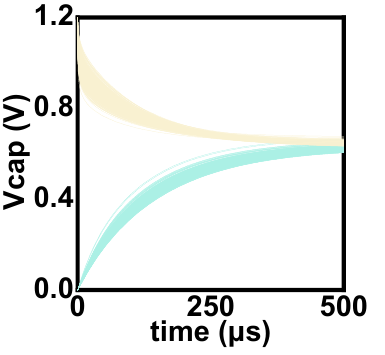}}
& \makecell{\includegraphics[width=2cm,height=2cm]{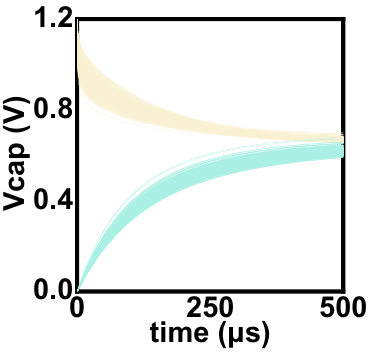}} 
& \makecell{\includegraphics[width=2cm,height=2cm]{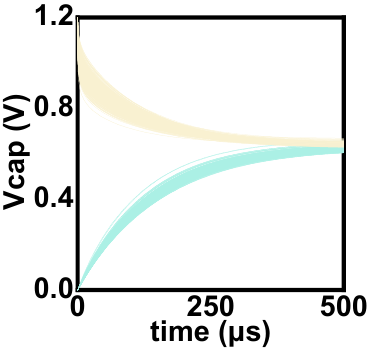}}
& \makecell{\includegraphics[width=2cm,height=2cm]{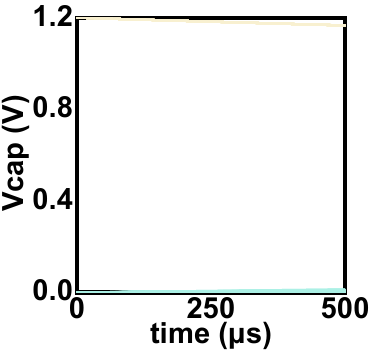}}
& \makecell{\includegraphics[width=2cm,height=2cm]{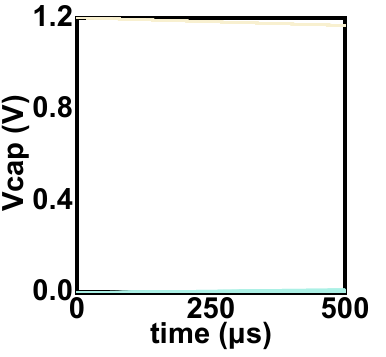}} \\
\hline
\end{tabular}
\end{table*}

The operation of the cell is as follows. As shown in the Fig. \ref{fig:dram_cell}(b), when the image sensor generates an event signal (EV) in response to intensity changes, it raises the Write Bit Line (WBL) voltage while simultaneously lowering the WWL voltage through an inverter. At this point, the LL switch is in a low-resistance state, allowing charge to flow into the storage capacitor ($C_{mem}$), which raises the memory cell voltage ($V_{mem}$) to $V_{dd}$. After the writing pulse, the switch turns off. However, charge leakage on $C_{mem}$ happens and $V_{mem}$ decreases near exponentially due to the leakage from the LL switch. This mechanism enables the normalization of timestamps directly within the memory unit.

In Fig. \ref{fig:dram_cell}(c) and (d), we compare the leakage characteristics of the LL switch and the conventional transmission gate (TG) in a SPICE simulation using a $65$ nm CMOS process. The leakage currents in a transistor can be classified into three main components: the channel leakage current ($I_{c}$), the body leakage current ($I_{b}$), and the gate leakage current ($I_{g}$) \cite{roy2003leakage}. $I_{c}$ flows through the channel between the source and drain, which is primarily the result of subthreshold conduction and further enhanced by the drain-induced barrier lowering (DIBL) effect. $I_{c}$ can be reduced exponentially by increasing the gate-to-source voltage ($V_{gs}$), reducing the drain-to-source voltage ($V_{ds}$), or by increasing the transistor's threshold voltage ($V_{th}$). $I_{b}$ mainly consists of reverse-biased parasitic diode leakage and gate-induced drain leakage (GIDL). $I_{b}$ can be mitigated by reducing the voltage differences between the transistor's terminals. $I_{g}$ is caused by tunneling and hot-carrier injection between the gate and channel, as well as edge direct tunneling in the overlap region between the gate and drain. In this work, we reduce $I_{g}$ by using thick oxide transistors. 

Based on the above analysis, we implemented a stacked structure with two PMOS transistors since it can halve the $V_{ds}$ across each transistor. During the initial leakage phase, $V_{mem}$ equals $V_{dd}$, and WBL is 0 V. In this configuration, the $V_{ds}$ of the left PMOS is approximately $\eta$ $V_{dd}$, while the $V_{ds}$ of the right PMOS is approximately ($1-\eta$) $V_{dd}$, where $\eta$ is a value between $0$ and $1$. Compared to the TG, the channel leakage current ($I_{c}$) in the off-state of the transistors is effectively minimized by reducing $V_{ds}$. By connecting the M-node to the well, the leakage current from the M-node to the well is also reduced. The well is floating similar to pseudoresistor designs in neural amplifiers \cite{reid_pseudo,sawan_pres} which require the creation of a large resistance to get very low cut-off frequencies. Increasing the channel length was found to have minimum impact and hence the size of the transistors were chosen based on the largest size that can fit under the capacitor, in order to reduce the variations. $C_{mem}$ is chosen based on the required temporal window as shown in Section \ref{subsec:results}. SPICE simulation results of Fig.\ref{fig:dram_cell}(d) show that, for $C_{mem}\approx20$ fF, the LL switch extends the effective retention time window to $>50$ ms. In contrast, the TG switch exhibits rapid charge leakage, with the charge completely dissipated in around $10$ ms.

\subsection{3D Architecture Design}

\begin{figure}[t]
    \centering
    \includegraphics[width=0.45\textwidth]{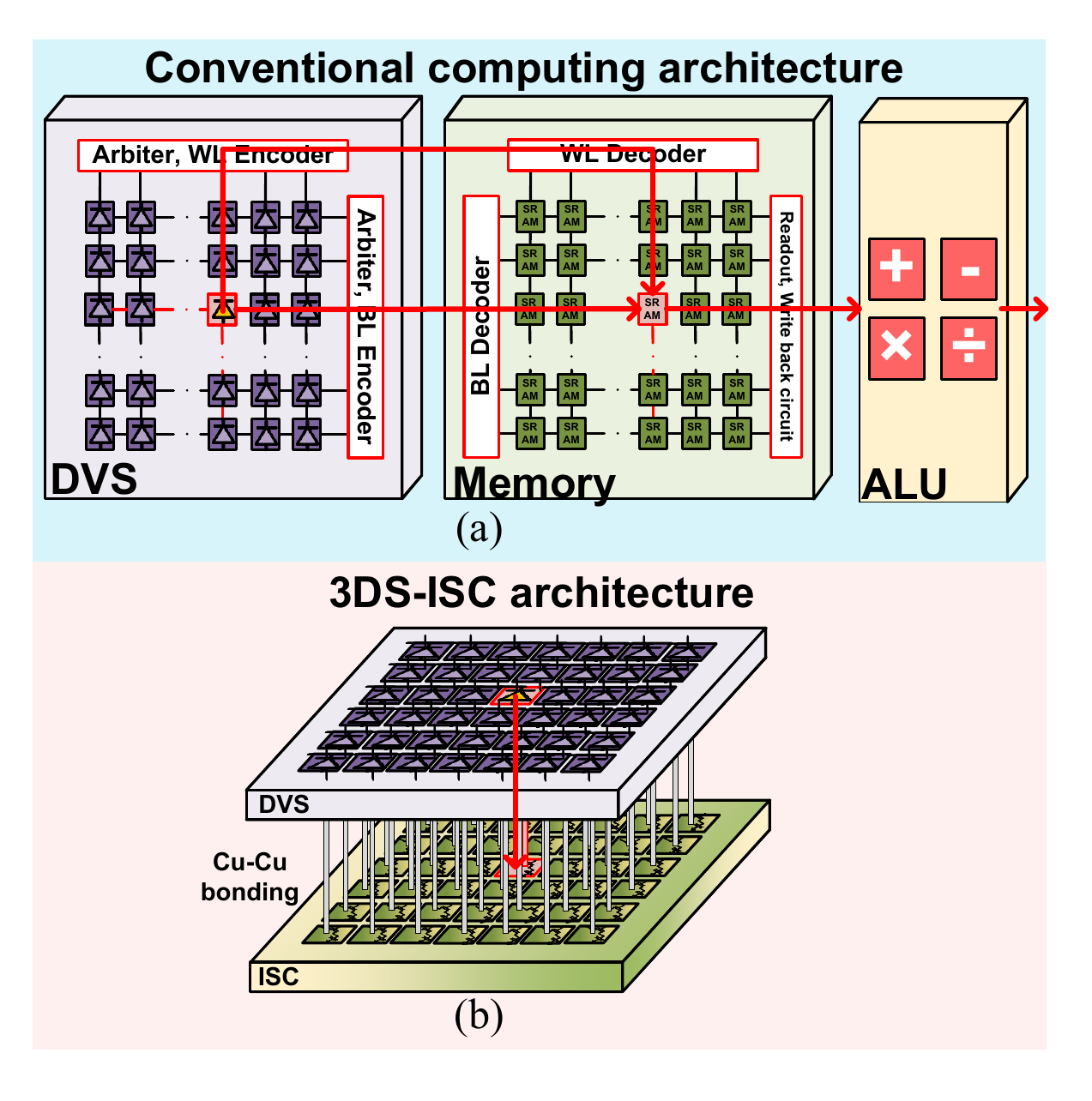}
    \caption{
    Comparison of 3DS-ISC and conventional architecture. The conventional 2D architecture requires charge and discharge of long wires spanning the entire array for each event while the 3D architecture has direct short connections between the sensor and eDRAM performing in-sensor computing.}
    \label{fig:threed_architecture} 
\end{figure}

Although event cameras are capable of extremely fast response times, their throughput is often limited by data transmission bandwidth (shown in Fig.\ref{fig:threed_architecture} (a)). In conventional 2D system architectures, every event from the sensor needs to pass through an AER encoder to create an address. This event along with the timestamp is then transferred to another memory to create a SAE. Finally, the memory transfers the data to computation units for normalization or further processing. This architecture introduces significant latency due to this long signal path.

To address these issues, we propose the 3D Stack In-Sensor Computing (3DS-ISC) architecture, leveraging 3D stacking technology to integrate the sensor and eDRAM-based ISC units into a vertically stacked chip. As shown in Fig.\ref{fig:threed_architecture} (b), events sensed by the DVS are directly written into the ISC cell by Cu-Cu bonding, and normalization operation is performed within the ISC unit as described earlier. This 3D architecture also helps to avoid a half-select issue where unintended voltage coupling can arise in a 2D implementation of this eDRAM based TS.

\begin{figure*}[htbp]
    \centering
    \includegraphics[width=0.85\textwidth]{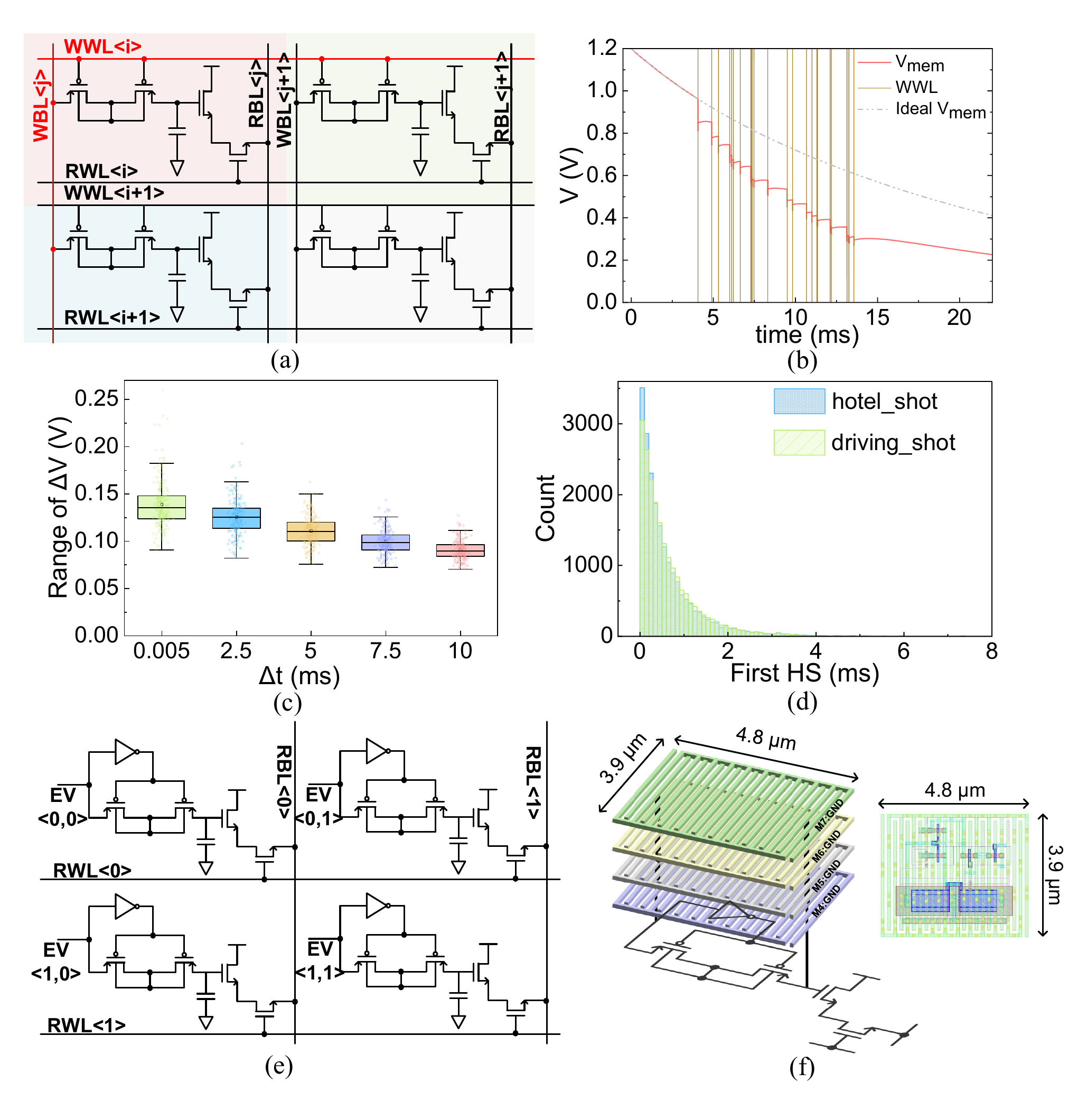}
    \caption{The Necessity of 3D Architecture Due to the Half-Select Problem.
     (a) The red background cell is fully selected, while cells with a blue and green background indicate that only the WBL is selected, and the WWL is selected, respectively. The cell with a gray background is unaffected by the half-select issue. (b) The impact of the green background half-select situation is significant. When the WWL is selected, there is a substantial decrease in the actual voltage $V_{mem}$ compared to the ideal voltage $V_{mem}$. (c) A Monte Carlo simulation is conducted to analyze the $\Delta$V (the voltage difference between actual $V_{mem}$ and ideal $V_{mem}$) based on the $\Delta$t (the time difference  between the write time and the half-select time) for an ISC cell. (d) The distribution of the first half-select time of the DND21 dataset. (e) The figure illustrates a 2$\times$2 structure of an ISC array. Each ISC unit operates independently thanks to the Cu-Cu bonding, in contrast to traditional storage arrays that use WWLs and WBLs.  (f) Under the TSMC 65nm process, the dimensions of a single ISC unit are 4.8 $\mu$m $\times$ 3.9 $\mu$m, with the minimum size being limited by $C_{mem}$. $C_{mem}$ consists of a metal-oxide-metal capacitor (MOMCAP) stacked from layers M4 to M7, and at this size, the MOMCAP value is 20 fF.}
    \label{fig:hs_issue} 
\end{figure*}

In the Fig.\ref{fig:hs_issue}(a), we illustrate different scenarios of cell selection during a write operation in a conventional 2D architecture with a crossbar type selection typical of 2D memory arrays. The red background cell is a fully selected one where an event is being written, where both the WWL$<$i$>$ and WBL$<$j$>$ are active, i.e. WBL is high while WWL is low. Half-selected cells are shown with green and blue backgrounds. For the green background half-selected cell, WWL$<$i$>$ is active, which means the LL switch is at the on state, but WBL$<$j+1$>$ is not selected and maintains at the low potential. In this case, the charge stored on the capacitor ($C_{mem}$) leaks through the LL switch into the WBL$<$j+1$>$, causing the voltage drop in the $V_{mem}$. The impact of the half-selection problem is analyzed in Fig.\ref{fig:hs_issue}(b) using events from a specific time segment from the DND21's hotel-bar dataset \cite{guo2022low}. The results show that the earlier the half-selection occurs after an event write (full selection), the greater the impact on $V_{mem}$. For the half-selected cell of blue background, where WWL$<$i+1$>$ is low while the WBL$<$j$>$ is high, the LL switch is off, but coupling capacitance between WBL$<$j$>$ and the PMOS's gate causes a small voltage fluctuation on $V_{mem}$. Gray background cell is the completely unselected cell, which are unaffected during the write operation.

To further investigate the relationship between the half-selection happening time ($\Delta$t) and the resulting degradation of $V_{mem}$ ($\Delta$V), we performed Monte Carlo simulations. As shown in Fig.\ref{fig:hs_issue}(c), earlier occurrences of half-selection after an event write (full selection) result in more significant $V_{mem}$ degradation. We also conducted a statistical analysis of the first half-selection time using the hotel-bar and driving datasets of DND21. The Fig.\ref{fig:hs_issue}(d) shows that the first occurrence of half-selection happens very early in both datasets. It can be observed that many half-select events occur relatively early, leading to a significant drop in the $V_{mem}$, which in turn results in considerable errors in the stored TS values. Therefore, when using a 2D structure that shares the WWL and WBL for both rows and columns, the half-select issue can cause substantial inaccuracies in the TS. In contrast, with a 3D structure, we can utilize Cu-Cu bonding to write events for each pixel \emph{individually without half-selection}, effectively avoiding this problem.


Fig.\ref{fig:hs_issue}(e) uses a 2$\times$2 array to show the array organization of the eDRAM cells. Each EV signal is connected to the Cu-Cu bonding and stimulated by the DVS pixel. This places stringent area restrictions on the eDRAM cell to fit below the DVS pixel. Fig.\ref{fig:hs_issue}(f) shows the layout of the eDRAM cell. The storage capacitor ($C_{mem}$) is implemented using a custom interdigitated capacitor structure, utilizing metal layers 4 to 7 for high-density storage. Each cell occupies $\approx 20\mu m^2$, which is smaller than most of existing DVS pixel sizes\cite{brandli2014240, posch2010qvga, serrano2013128}. The size of the ISC cell is limited by the size of this MOMCAP. In this layout size, $\approx 20$ fF capacitance can be obtained. The next section will show with Monte Carlo simulations that this size enables sufficient matching between cells. Another potential architecture has one DVS pixel for every $n\times n$ block of normal image sensor pixels where the DVS and sensor circuits are 3D stacked in a layer below the photodiodes \cite{guo2023three}. In that case, the eDRAM cells for the TS generation can be placed in the same layer and next to the DVS pixel.


\section{Results}
\label{subsec:results}
In this section, we present circuit and algorithm simulation results of our proposed architecture. 
\subsection{Characterization of ISC layer}
\label{subsec:characterize}

We first characterize the properties of the eDRAM cells in the ISC layer designed in a $65$ nm CMOS process. By comparing leakage characteristics and presenting array-level simulation results, we demonstrate the robustness of the proposed circuit design. 
\begin{figure}[htbp]
    \centering
    \includegraphics[width=0.48\textwidth]{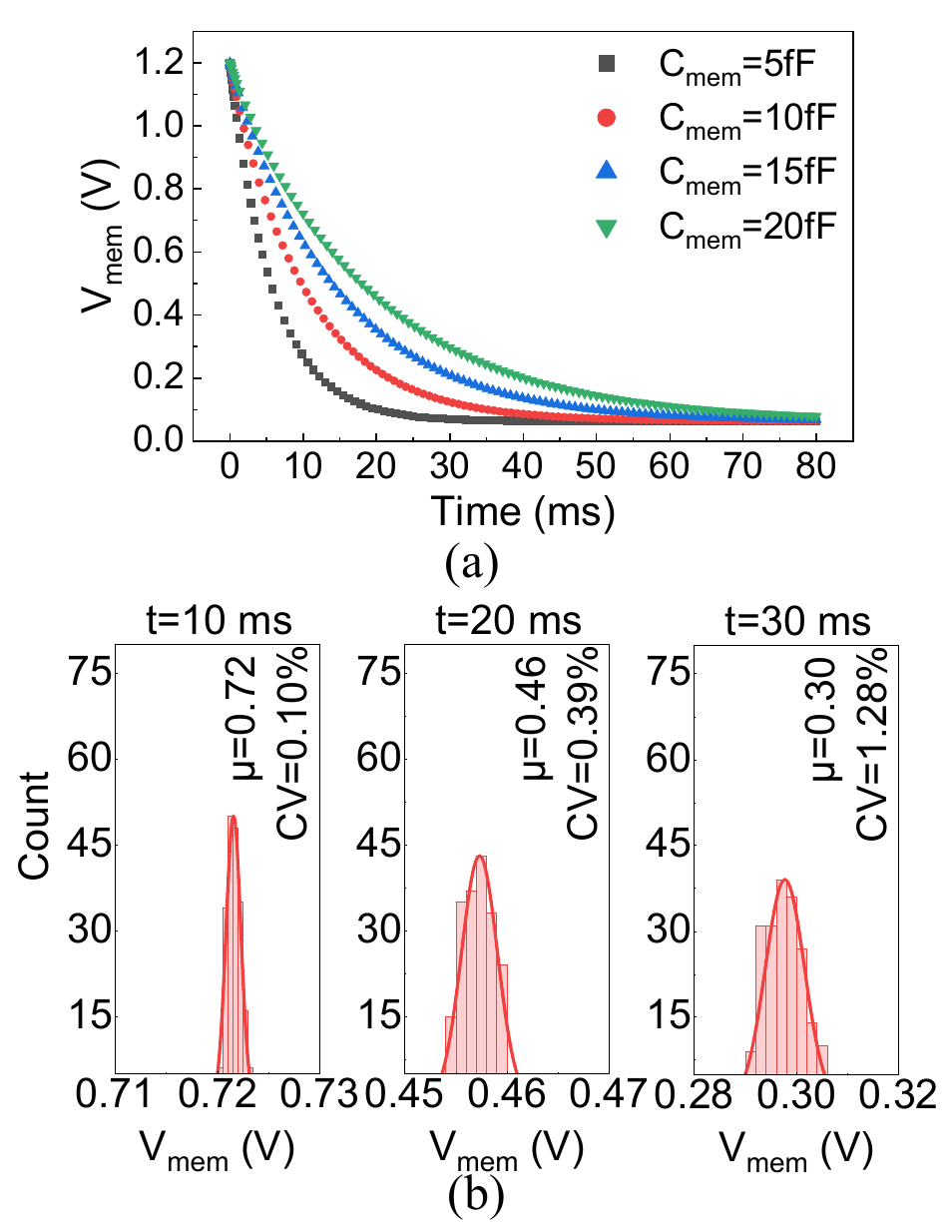}
    \caption{
     Memory characterization of eDRAM cells. (a) Comparison of the voltage decay of $V_{mem}$ for various $C_{mem}$ in the eDRAM cells. $C_{mem}\geq 10$ fF is necessary to achieve a memory of at least $24$ ms. (b) For the case of $C_{mem}=20$ fF, a Monte Carlo simulation of $V_{mem}$ at different times after the event write shows a coefficient of variation $<2$\%.}
    \label{fig:change_cap} 
\end{figure}

\begin{figure}[t]
    \centering
    \includegraphics[scale=0.5]{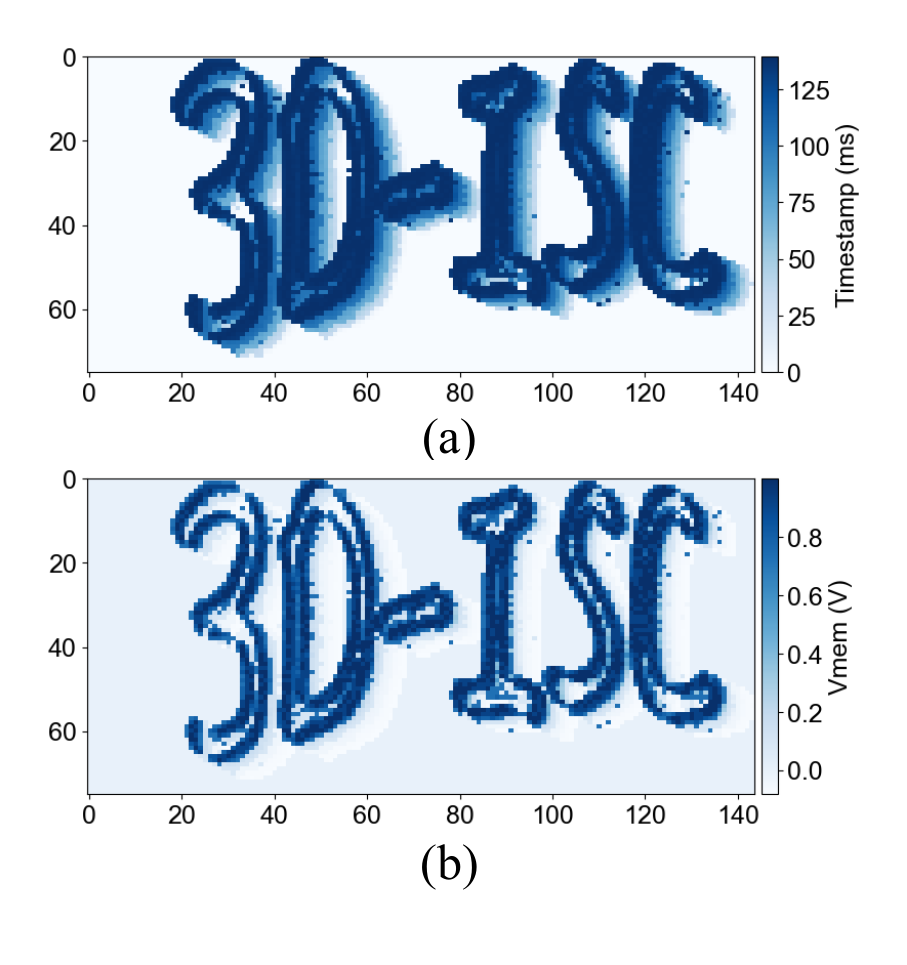}
    \caption{Visualization of array-level simulation. (a) The timestamps of a particular event sequence and (b) the values of $V_{mem}$ to create the proposed analog TS.}
    \label{fig:TS_qualitative} 
\end{figure}

Fig.\ref{fig:change_cap}(a) demonstrates the impact of the size of $C_{mem}$ on the retention time or memory window. As expected, increasing the value of $C_{mem}$ significantly extends the charge retention time at $V_{mem}$, effectively enlarging the memory window. This is because a larger capacitor can store more charge. By carefully selecting the value of $C_{mem}$, a balance can be achieved between extending the time window and minimizing the area overhead. As shown later and in other works\cite{guo2022low}, algorithmic requirements need a memory window $\geq 24$ ms necessitating a $C_{mem}\geq10$ fF based on this simulation. Another factor in choosing the capacitor size is cell to cell variability. This is also affected by mismatch in the leakage current of the pseudo resistor which is reduced by choosing the largest PMOS sizes that can fit beneath the capacitor. Fig.\ref{fig:change_cap}(b) presents Monte Carlo simulation results for a $C_{mem}=20$ fF at different times of the decay waveform, i.e. at different delays $\Delta t$ after the event write. At $\Delta t=10$ ms, the average value of $V_{mem}$, $\mu=0.72$ V with coefficient of variance $CV = 0.10\%$. At $\Delta t = 20$ ms, $\mu = 0.46$ V with $CV = 0.39\%$. Finally, at $\Delta t = 30$ ms, $\mu = 0.30$ V and $CV = 1.28\%$. The results reveal that the distribution is reasonably concentrated with coefficient of variation $< 2\%$. As shown in the next subsection, this variability does not degrade the quality of the TS to affect the performance in various algorithms. Here, we show a qualitative view of the hardware TS with simulated variability in Fig. \ref{fig:TS_qualitative}. The top panel depicts the SAE, i.e. it directly displays the timestamps written into the array. The bottom panel shows the TS represented by $V_{mem}$ of the ISC array including the simulated cell to cell variability. The latest event corresponds to a $V_{mem}$ closer to $1$ V, while the older events correspond to values closer to $0$ V.

\subsection{Advantages of 3D architecture and eDRAM ISC unit}
\label{subsec:compare}

Both the 3D stacked architecture and the eDRAM ISC analog circuit provide significant advantages in improving the system performance. To investigate these advantages, we have conducted a comparative analysis in two aspects: (1) the benefits of 3D stacking architecture versus traditional 2D architectures, and (2) the performance improvements brought by analog ISC circuit over conventional digital processing with Static Random Access Memory (SRAM) to store the timestamps. Both of these two comparisons are made under QVGA resolution (320$\times$240).

In the evaluation process, detailed calculation and analysis are carried out for the different circuit modules. In order to fully evaluate the power consumption characteristics of the circuit, we calculated the static power consumption and the dynamic power consumption separately. Static power consumption is mainly caused by the leakage current. Dynamic power consumption is caused by the switching activity of the circuit, and related to the frequency of the event arrival. We use an event frequency of $100$ Meps that is representative of modern DVS\cite{gallego2020event}. 
For the power consumption related to Cu-Cu bonding, we refer to the model proposed in \cite{ku2018area}. The power consumption is due to the parasitic capacitor and resistance of the Cu-Cu bonding which are $0.5$ fF and $0.2$ $\Omega$ respectively as mentioned in\cite{ku2018area}. For comparison with digital processing where time stamps are recorded using SRAM, memory specifications were obtained from \cite{bose202151} and \cite{rios2023within}. In \cite{bose202151}, 5.1 pJ is consumed for writing one bit in SRAM, and static leakage current for SRAM is 350 pA for $1$ V power supply. For \cite{rios2023within}, the static power consumption of an SRAM array storing 346×260 pixels, with each pixel using 18 bits, is 35 mW. The power consumption for accessing a 7×7 pixel SRAM is 2.4 nJ. The write power consumption is approximately 1.5-6$\times$ that of the read power consumption\cite{4140596,bose202151}; we choose a conservative estimate of $1.5\times$ in this analysis. Based on these data, we can estimate both the static and dynamic power consumption of their designs operating at QVGA resolution with a $16$-bit timestamp. The power assessment of the ISC array is obtained by Cadence Virtuoso. For the power assessment of other peripheral circuit modules (such as decoders, etc.) in the system, Synopsys DC Compiler is used for comprehensive analysis. 

\begin{figure}[t]
    \centering
    \includegraphics[width=0.5\textwidth]{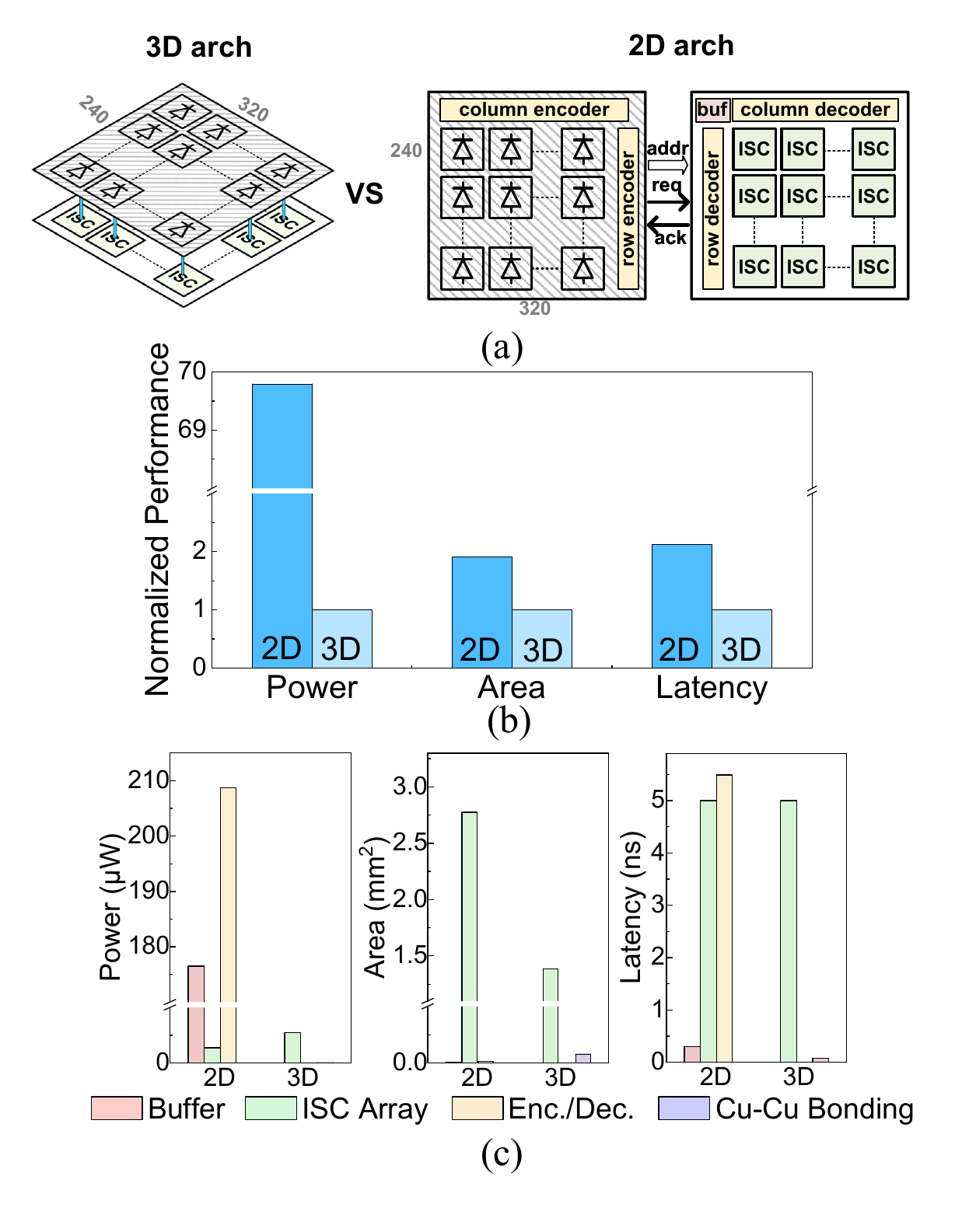}
    \caption{A comparison between the 3D architecture and 2D architecture for a $320\times 240$ array under 100 MHz. (a) The architecture of 3D and 2D implementation. (b) The 2D architecture does not use Cu-Cu bondings compared to the 3D architecture, but adds encoders, decoders, and buffers for driving the longer WWLs and WBLs. The power consumption, area, and delay of both architectures are compared, (c) with a breakdown of the contributions from each module.}
    \label{fig:2d_3d_comparison} 
\end{figure}

In Fig. \ref{fig:2d_3d_comparison}, the performance of the 3D and 2D architectures is compared using the same eDRAM ISC unit across three key metrics: power consumption, area overhead, and transmission delay. The results demonstrate that the 3D architecture achieves significant improvements over its 2D counterpart, with a 69$\times$ reduction in power consumption, a 1.9$\times$ reduction in area, and a 2.2$\times$ reduction in delay. These enhancements can be attributed to the inherent advantages of the 3D stacked architecture. For the 2D architecture, the design must account for the decoders and encoders, as well as buffers that drive long metal wires (WBLs and WWLs). While Cu-Cu bondings also introduce power, area and delay overhead due to data transmission, these costs are far outweighed by its benefits as analyzed in details below. 


From the power breakdown in Fig. \ref{fig:2d_3d_comparison}(c), the 2D architecture is dominated by two contributors: the encoder/decoder behavior (53.8\% of the total) and the buffers that charge and discharge the WWL and WBL (45.5\%). These components are reduced in the 3D-ISC architecture. Instead, the power of 3D-ISC is primarily driven by ISC-array activity. In the area breakdown, the 3D-ISC array occupies roughly half the area of its 2D counterpart owing to compact vertical stacking. The additional buffers and encoder/decoder blocks required by the 2D design make up only a small fraction of the total area, and the Cu–Cu bonding footprint is also minimal. In terms of latency, 3D-ISC reduces the total latency from $\sim$11 ns to $\sim$5 ns. Both architectures exhibit similar event-write latency ($\sim$5 ns). The 2D architecture further incurs $\sim$6 ns due to encoder/decoder and handshaking overhead \cite{chen2021development}, collectively contributing 46.4\% of its total latency. By contrast, the Cu–Cu bonding latency in 3D-ISC is only $\sim$0.08 ns \cite{ku2018area}, which is negligible and underscores its efficiency in data movement.

\begin{figure}[t]
    \centering
    \includegraphics[width=0.45\textwidth]{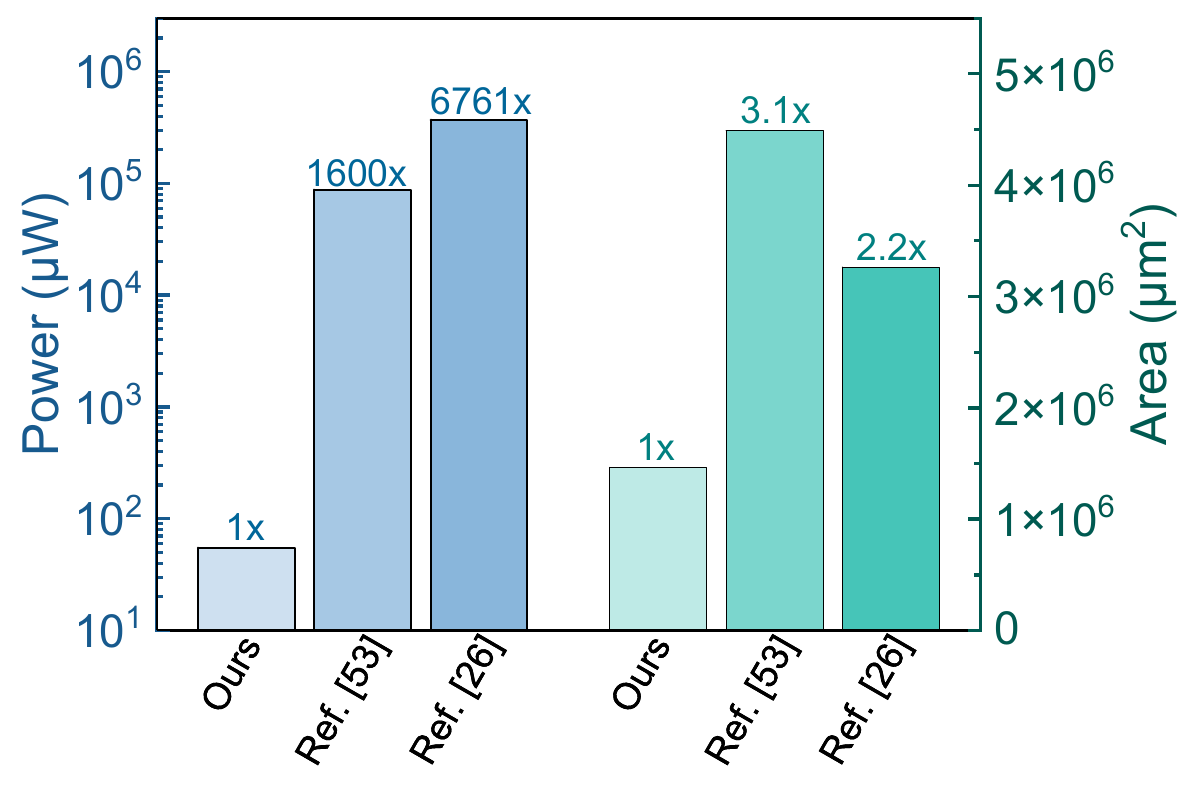}
    \caption{A comparison of ISC analog array and SRAM implementations\cite{bose202151,rios2023within}. The analog eDRAM based TS is 1600$\times$ and 6761$\times$ more power efficient. The area required by our eDRAM cell is 3.1$\times$ and 2.2$\times$ less than the area needed by the $16$-bit SRAM cells to store digital timestamps.}
    \label{fig:compare_sram} 
\end{figure}

In comparing the ISC analog implementation with the SRAM digital implementations, we focus exclusively on the storage array. As shown in Fig.\ref{fig:compare_sram}, our ISC array demonstrates significantly lower power and area consumption compared to the two kinds of SRAM implementations. \cite{bose202151} consumes 1600$\times$ power and occupies 3.1$\times$ area. \cite{rios2023within} shows 6761$\times$ power and 2.2$\times$ area. These results demonstrate the superior efficiency of our design in both power and area when compared to the existing SRAM methods.

Furthermore, the SRAM implementation suffers from periodic errors due to the timestamp overflow issue. In contrast, our eDRAM array effectively avoids this challenge thanks to its self-normalization property.

\subsection{Application 1: Implementation of Denoise algorithm on 3DS-ISC}
\label{subsec:denoise}

\begin{figure}[htbp]
    \centering
    \includegraphics[width=0.5\textwidth]{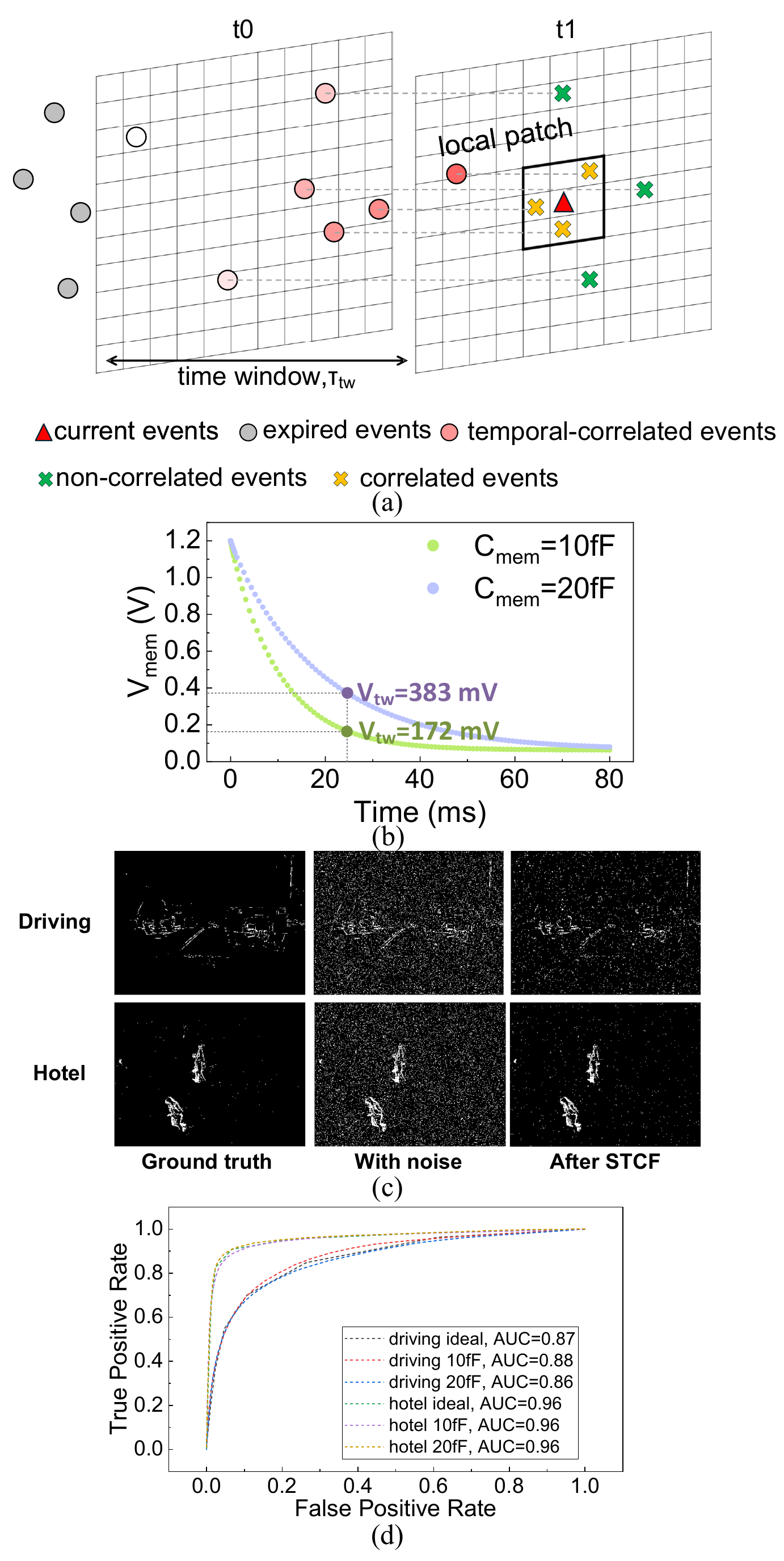}
    \caption{ Denoising Effects of the STCF Algorithm Combined with 3DS-ISC. 
     (a) The STCF algorithm utilizes the temporal and spatial correlations of events to denoise signals. If the number of correlated events exceeds a certain threshold, the event is deemed valid. (b) The correspondence between the time window and $V_{mem}$. When the time window is 24 ms, the corresponding $V_{mem}$ values are 383 mV and 172 mV for 20 fF and 10 fF, respectively. (c) Visualization of denoise effects on the ``driving" and ``hotel-bar" datasets \cite{guo2022low}. (d) Binary classification ROC curve where "ideal" represent software results, while ``10 fF/20 fF" refers to the application of STCF based on the TS constructed by the ISC layer, under the conditions of $C_{mem}$ being 10 fF and 20 fF. Both 10 fF and 20 fF choices result in acceptable AUC values. }
    \label{fig:STCF} 
\end{figure}

We first evaluate the accuracy of our approach on the important pre-processing step of noise removal that uses the TS. Spatial-temporal correlation filter (STCF)\cite{guo2022low} is a denoise filter for DVS, allowing only events that exhibit both spatial correlation and temporal correlation to pass through. As shown in Fig.\ref{fig:STCF}(a), the red triangle appearing at time $t_1$ represents the current detected event that needs to be classified. The spatial correlation is defined by the local patch which surrounds the current event. The time correlation is defined by a time window ${\tau}_{tw}$, from the present time to the past. The circular markers depict events from past moments, with colors closer to red indicating proximity to the present time, while those closer to white suggest events further in the past. Grey markers represent events that have exceeded ${\tau}_{tw}$, losing temporal relevance with the current time. The 'X' symbols denote mappings of temporally correlated past events at t1. A yellow 'X' within the local patch indicates spatial relevance, whereas a green 'X' signifies events beyond the local patch. Hence, supporting events showing both temporal and spatial relevance are depicted as yellow 'X'. In this example, there are three valid historical events. Then, if the number of supporting events is higher than a defined threshold $th$, the current event is classified as a signal event; otherwise, it is identified as a noise event. In our eDRAM-based ISC architecture, the value of $V_{mem}$ represents the temporal proximity to t1. When $V_{mem}$ is greater than a certain voltage $V_{tw}$, it indicates that the timestamp stored in that pixel lies within the ${\tau}_{tw}$ (see Fig. \ref{fig:STCF}(b)). Otherwise, it means that the event occurred too long ago and is outside the ${\tau}_{tw}$. ${\tau}_{tw}$ is chosen as 24 ms here. A simple comparator can be used in the post-processing stage to compare $V_{mem}$ with $V_{tw}$.

\begin{figure}[t]
    \centering
    \includegraphics[width=0.3\textwidth]{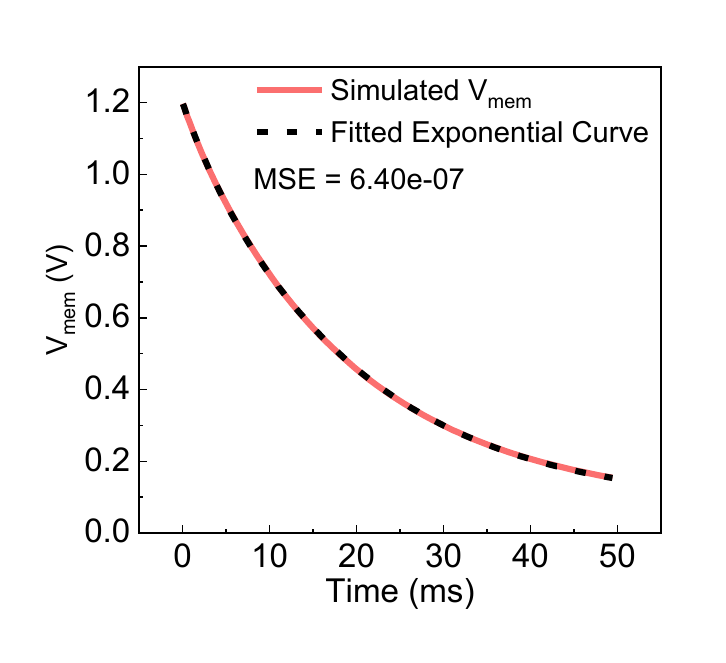}
    \caption{Decay of $V_{mem}$ with time in SPICE simulation and the fitted exponential curve.}
    \label{fig:mse} 
\end{figure}

It is impractical to perform SPICE simulation for entire datasets due to the prohibitive computational complexity and time required. Therefore, we made a computational model based on SPICE simulations. This allows us to incorporate circuit non-idealities into subsequent analysis and modeling. First, for modeling the charge loss at each pixel, we employed the following normalized exponential function: $f(t) = A_1\exp\left(-\frac{t}{\tau_1}\right)+A_2\exp\left(-\frac{t}{\tau_2}\right)+b$. As shown in Fig. \ref{fig:mse}, the mean squared error (MSE) between the simulated $V_{mem}$ and the fitted exponential curve indicates a very good fit. Next, we obtained the circuit non-uniformity or pixel to pixel variability through Monte Carlo simulations. Pixel variability in the software model was generated by randomly sampling from 8,000 MC simulations, each fitted to the double-exponential decay function, with the corresponding parameters mapped to individual pixels.

As shown in Fig. \ref{fig:STCF}(c), we used the above method to emulate the performance of our eDRAM ISC approach on the DND21 dataset \cite{guo2022low}, comprising two classes: hotel-bar and driving. The hotel-bar dataset was captured using a DAVIS346 camera, recording under stationary camera conditions. On the other hand, the driving dataset represents a simulated scenario of a car moving through a city, converted from video to events through the tool v2e\cite{hu2021v2e}. The ground truth denotes the clean dataset, containing only real event results, while the one with noise indicates the addition of 5 Hz/pixel noise to these datasets following the approach in \cite{guo2022low}. The results of STCF filter are obtained by using the hardware emulation described earlier. All images are records of events over $20$ ms. It is evident that our circuit design is effective for the STCF algorithm. For quantitative analysis, the ROC curves are drawn, comparing the ideal results (which use full-precision timestamp to construct TS) and our ISC hardware implementation results (which use $V_{mem}$ to represent the TS). Both the values of 10 fF and 20 fF are considered for $C_{mem}$. Experimental results in Fig. \ref{fig:STCF}(d) show that for both datasets, either value of capacitance is acceptable for noise filtering. This implies that there is scope to reduce the bit-cell area if the TS is used for simple tasks like denoise.

\subsection{Application 2: Implementation of Image classification on 3DS-ISC}
\label{subsec:classify}

\begin{figure*}[t]
    \centering
    \includegraphics[width=0.7\textwidth]{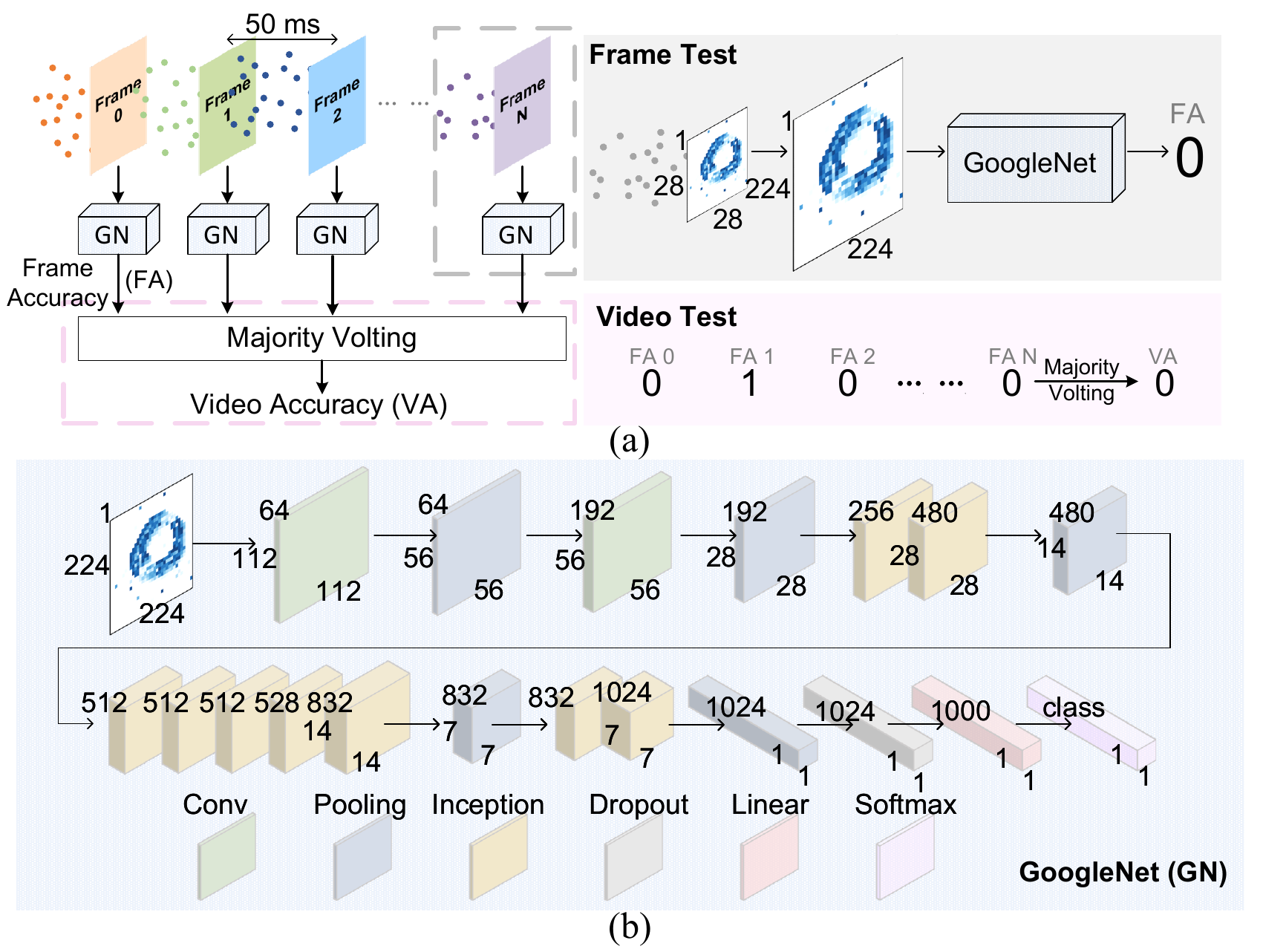}
    \caption{The event stream is converted into a TS and classified using GoogleNet for frame accuracy and using the majority voting method for video accuracy. (a) For frame test, every 50 milliseconds of events, the events are accumulated. Utilizing the simulated hardware performance of the 3D-ISC, a frame of the TS is generated. Given the varying sizes of the dataset images, interpolation is employed to scale the TS to a size of 224x224. The TS is then used as input for classification by GoogleNet to obtain the frame accuracy. To obtain one classification label per sample, the video accuracy is evaluated, which is found by the majority voting \cite{mohan2022ebbinnot,ussa2023hybrid} across the frame accuracies in the video. (b) The network structure of GoogleNet is employed for the CNN. }
    \label{fig:googlenet} 
\end{figure*}

The TS constructed by 3DS-ISC can be processed using traditional CV algorithms. To evaluate its performance in classification tasks, we treated the TS as a two-dimensional image and input it into a convolutional neural network (CNN). 
In this work, the GoogLeNet architecture \cite{szegedy2015going} was employed and initialized with ImageNet-pretrained weights. During the training of GoogLeNet, the optimizer used was Adam, with an initial learning rate of 0.0001 and default momentum parameters betas=(0.9, 0.999). 
The training was conducted for a total of 20 epochs. 

Experiments were conducted on various widely used event camera datasets, including N-Caltech101 \cite{orchard2015converting}, N-MNIST\cite{orchard2015converting}, CIFAR10-DVS\cite{li2017cifar10} and DVS128 Gesture\cite{amir2017low} . The datasets were all split to training and test sets for model fine-tuning and evaluation. The N-MNIST and DvsGesture were split according to official train/test labels, while the others were randomly split in the ratio of 8: 2. For each video in the datasets, every $50$ ms events were extracted to construct the TS and used as input for the model. In addition, the input TS was resized to 224$\times$224, due to the limit of GoogLeNet. Depending on the duration of each video, a varying number of frames were generated. Both frame accuracy and video accuracy \cite{mohan2022ebbinnot, ussa2023hybrid} were evaluated. The video accuracy was determined by majority voting \cite{mohan2022ebbinnot, ussa2023hybrid} over all frames within a sample, thereby fully leveraging the temporal information within each sample. This results in one output per sample which makes it a fair comparison with other methods.

\begin{table*}[ht]
    \centering
    \caption{Classification Accuracy Compared to Other Methods on Public Datasets. Accuracy reported as frame accuracy/video accuracy.}
    \label{tab:googlenet}
    \begin{tabular}{cccccc} 
        \toprule
        Method                 & Classifier &N-MNIST\cite{orchard2015converting}  & N-Caltech101\cite{orchard2015converting} & CIFAR10-DVS\cite{li2017cifar10} & DVS128 Gesture\cite{amir2017low} \\ 
        \midrule 
        HOTS\cite{lagorce2016hots}& SVM        & 0.81       & 0.21          & 0.27          & -      \\
        HATS\cite{sironi2018hats} & SVM        & 0.99       & 0.64          & 0.52          & -      \\
        H-First\cite{orchard2015hfirst}&SNN    & 0.71       & 0.05          & 0.08          & -      \\
        Gabor \cite{lee2016training}&SNN       & 0.84       & 0.20          & 0.25          & -      \\
        E2VID\cite{rebecq2019events}& CNN      & 0.98       & 0.87          & -             & -      \\
        SAE\cite{benosman2013event}& CNN       & 0.99       & 0.65          & -             & 0.95   \\
        TORE\cite{baldwin2022time}& CNN        & 0.99       & 0.83          & -             & 0.96   \\
        3DS-ISC(ours)             & CNN        & 0.99/0.99 & 0.82/0.85    & 0.72/0.78    & 0.91/0.97  \\
        \bottomrule 
        
    \end{tabular}
\end{table*}

The experimental results are shown in Table \ref{tab:googlenet}. For the frame accuracy, the 3D-ISC method achieved a classification accuracy of 99\% on the N-MNIST dataset and 72\% on the CIFAR10-DVS dataset, comparable to the best achievable performance using TS representation. On the N-Caltech101 dataset, our method achieved an accuracy of 82\%, outperforming most of other methods, including HOTS (21\%), HATS (64\%), Gabor (20\%), TORE (83\%), etc. On the DVS128 Gesture dataset, our method also showed a comparable results with SAE and TORE method. However, the hardware efficiency of our method is much better than SRAM-based implementation of SAE as shown in Sec. \ref{subsec:compare}. TORE\cite{baldwin2022time} requires FIFOs of depth $k$ (typically $\geq$ 3) per pixel storing 16-32 bit floating point numbers with two polarities. Hence, it requires at least $96$-bit FIFO per pixel resulting in 16 times more area than our approach. 
For video accuracy, the accuracy is further improved. Except for the results for the N-Caltech101 dataset (slightly worse than the E2VID method), our proposed method achieves the best accuracy for all other datasets. These results demonstrate that the TS generated by 3DS-ISC can effectively capture the temporal information of events.

\subsection{Application 3: Implementation of Image reconstruction on 3DS-ISC}

\begin{table}[t]
\centering
\caption{Image reconstruction results on DAVIS datasets compared with prior works}
\label{tab:image_reconstruction}
\begin{tabularx}{\linewidth}{l *{3}{>{\centering\arraybackslash}X}}
\toprule
\multirow{2}{*}{\textbf{Dataset}} & \multicolumn{3}{c}{\textbf{SSIM}} \\
\cmidrule(lr){2-4}
                                  & E2VID & TORE & 3D-ISC \\
\midrule
boxes\_6dof      & \textbf{0.63} & 0.53 & 0.41 \\
calibration      & 0.52 & 0.52 & \textbf{0.65} \\
dynamic\_6dof    & 0.50 & \textbf{0.55} & \textbf{0.74} \\
office\_zigzag   & 0.50 & 0.47 & \textbf{0.55} \\
poster\_6dof     & \textbf{0.68} & 0.59 & 0.67 \\
shapes\_6dof     & 0.44 & 0.63 & \textbf{0.91} \\
slider\_depth    & \textbf{0.61} & 0.59 & 0.40 \\
\midrule
mean             & 0.56 & 0.55 & \textbf{0.62} \\
\bottomrule
\end{tabularx}
\end{table}

Reconstructing intensity frames from events is a canonical task in EBCs, allowing the generation of images or video sequences. Such reconstruction can be achieved at high frame rates, offering high-speed video with minimal data bandwidth requirements. The DAVIS240C dataset\cite{mueggler2017event} provides paired event streams and APS frames, which we use for supervised image reconstruction. APS frames serve as ground-truth targets. Unlike the image classification approach, where frames are typically divided every 50 ms, the corresponding event streams are segmented based on APS frame timestamps to ensure precise temporal alignment. For each frame, the events are used to construct a $256 \times 256$ single-channel grayscale image from 3D-ISC. The resulting TS is then fed into the UNet model, with APS frames providing supervision, enabling end-to-end training for high-quality event-based image reconstruction. The network is trained for 50 epochs using an AdamW optimizer, with the learning rate adjusted by a "ReduceLROnPlateau" scheduler.

The quantitative results of image reconstruction are summarized in Table \ref{tab:image_reconstruction}, which reports Structure Similarity Index Measure (SSIM) scores on several DAVIS240C datasets and compares our method with E2VID\cite{rebecq2019events} and TORE \cite{baldwin2022time}. Overall, the proposed approach (3D-ISC) achieves the best average SSIM (0.62), outperforming both E2VID (0.56) and TORE (0.55). Notably, our method yields substantial improvements on challenging sequences such as dynamic\_6dof and shapes\_6dof, where it reaches 0.74 and 0.91, respectively. These results demonstrate that TS given by 3DS-ISC training with APS supervision significantly enhances the structural fidelity of the reconstructed images across diverse motion patterns and scenes.

\subsection{Discussion: Polarity Sensitive TS}
\begin{figure}[t]
    \centering
    \includegraphics[width=0.45\textwidth]{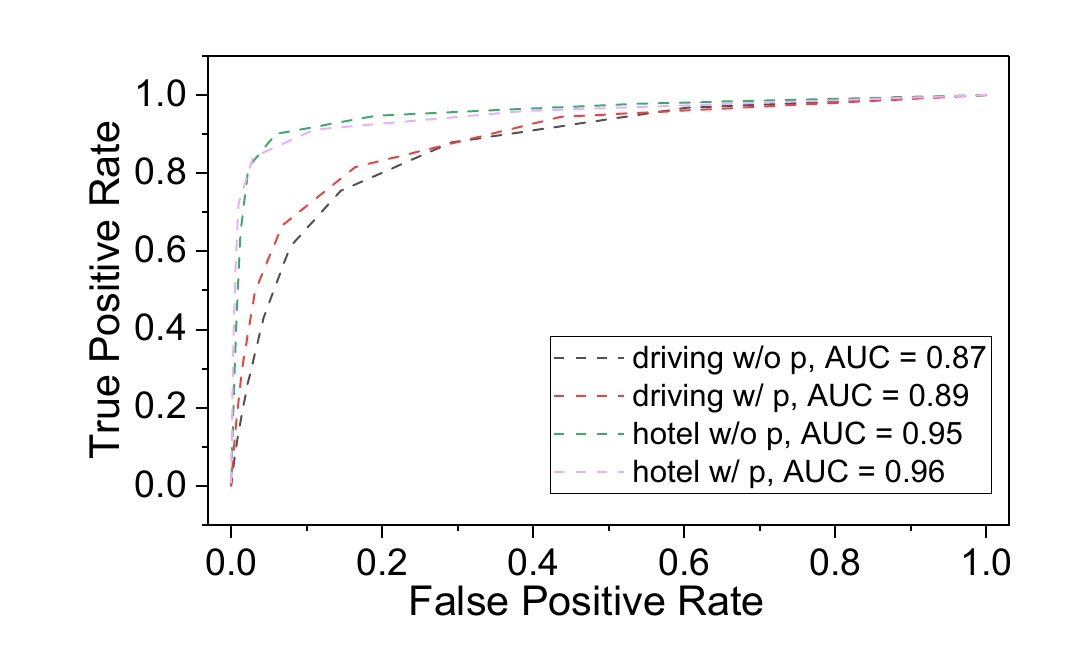}
    \caption{Results of the STCF algorithm with and without considering polarity. From the ROC curve, it can be seen that the impact of polarity on the results is minimal.}
    \label{fig:polarity_on_debnoise} 
\end{figure}

In the earlier results, we ignored event polarity, but it may be useful in some applications. Our proposed approach can be used to store separate TS for each event polarity, albeit at the cost of $2 \times$ the area. We evaluated the potential gains of this approach for the denoise algorithm. As shown in Fig.\ref{fig:polarity_on_debnoise}, when applying the denoising algorithm in Section IV.C with polarity, the AUC increases by only 2\% in the driving data set and 1\% in the hotel-bar data set. Therefore, the polarity can be ignored to some extent for this application. However, we found that for more complex tasks such as classification of CIFAR10-DVS, the polarity information does play a crucial role. When processing events with separated polarities, we achieved frame accuracy of 74\% and video accuracy of 80\%, representing substantial improvements over single-polarity processing reported in Table \ref{tab:googlenet}. This demonstrates that while polarity may be negligible for simpler tasks, it can enhance the performance of complex tasks.

\section{Conclusion} 

This study presents a comprehensive algorithm, architecture, and circuit co-design for TS construction, using a 3D stacked array of 6T-1C eDRAM cells. Our proposed 3DS-ISC architecture demonstrate significant advantages in terms of the computational efficiency, area utilization, and real-time performance. 

The results highlight the strengths of the proposed 3DS-ISC architecture and ISC circuit. By leveraging 3D stacking technology with Cu-Cu bondings, we achieved substantial improvements over traditional 2D designs in terms of power consumption (69$\times$), area (1.9$\times$), and delay (2.2$\times$). 
The 3D cell also removes a half-selection issue that occurs in the 2D architecture. In comparison to conventional SRAM-based timestamp storage, our ISC array offers significant improvements in the power consumption (1600-6761$\times$), area overhead (2.2-3.1$\times$) and avoids the data overflow problem. 

To test the utility of the TS in algorithms, we evaluated its performance on three tasks. The STCF algorithm successfully filtered noise from event datasets, achieving an AUC of 0.86 and 0.96 respectively on the driving and hotel-bar datasets. For classification tasks, our system achieved competitive  accuracy across multiple datasets, including 99\% on N-MNIST, 85\% on N-Caltech101, 78\% on CIFAR10-DVS and 97\% on DVS128 Gesture. Furthermore, 3D-ISC method achieves the highest average SSIM (0.62) in image reconstruction, significantly outperforming E2VID (0.56) and TORE (0.55), with substantial improvements on challenging sequences. In the future, integrating post-processing computational circuits directly into the unit presents a promising research direction.

%

\bibliographystyle{IEEEtran}
\bibliography{ref}

\begin{IEEEbiography}[{\includegraphics[width=1in,height=1.25in,clip,keepaspectratio]{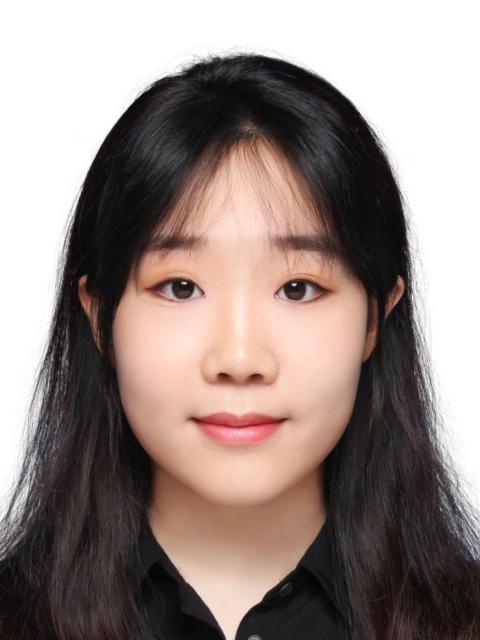}}]{Hongyang Shang} received her B.S. degree in electronic science and technology from Nankai University, and M.S. degree in electronics from Nanyang Technology University, respectively. After that, she worked as a research assistant in Fudan University. She is currently a PhD student in the Department of Electrical Engineering at City University of Hong Kong. Her research interests include bio-inspired neuromorphic circuits, event-based camera and Computing-In-Memory.
\end{IEEEbiography}

\begin{IEEEbiography}
[{\includegraphics[width=1in,height=1.25in,clip,keepaspectratio]{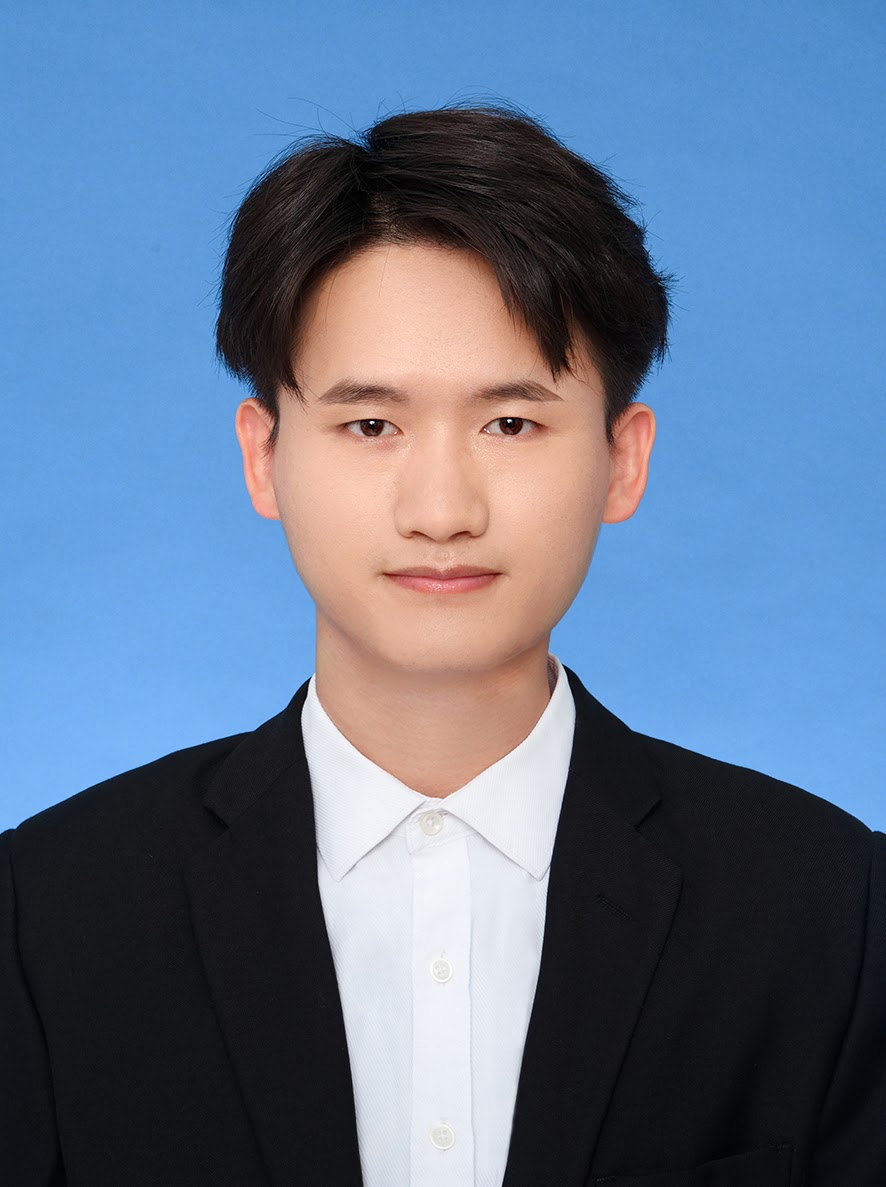}}]{Shuai Dong} received his B.S. degree in Electronic and Information Engineering from Chongqing University and M.Sc in Microelectronics and Solid-State Electronics from South China Normal University. He is currently a PhD student at City University of Hong Kong. His research interests include the development of in-memory computing architectures based on SRAM and RRAM, with a focus on low-power design, high-density integration, and efficient hardware implementation of neural networks.
\end{IEEEbiography}

\begin{IEEEbiography}
[{\includegraphics[width=1in,height=1.25in,clip,keepaspectratio]{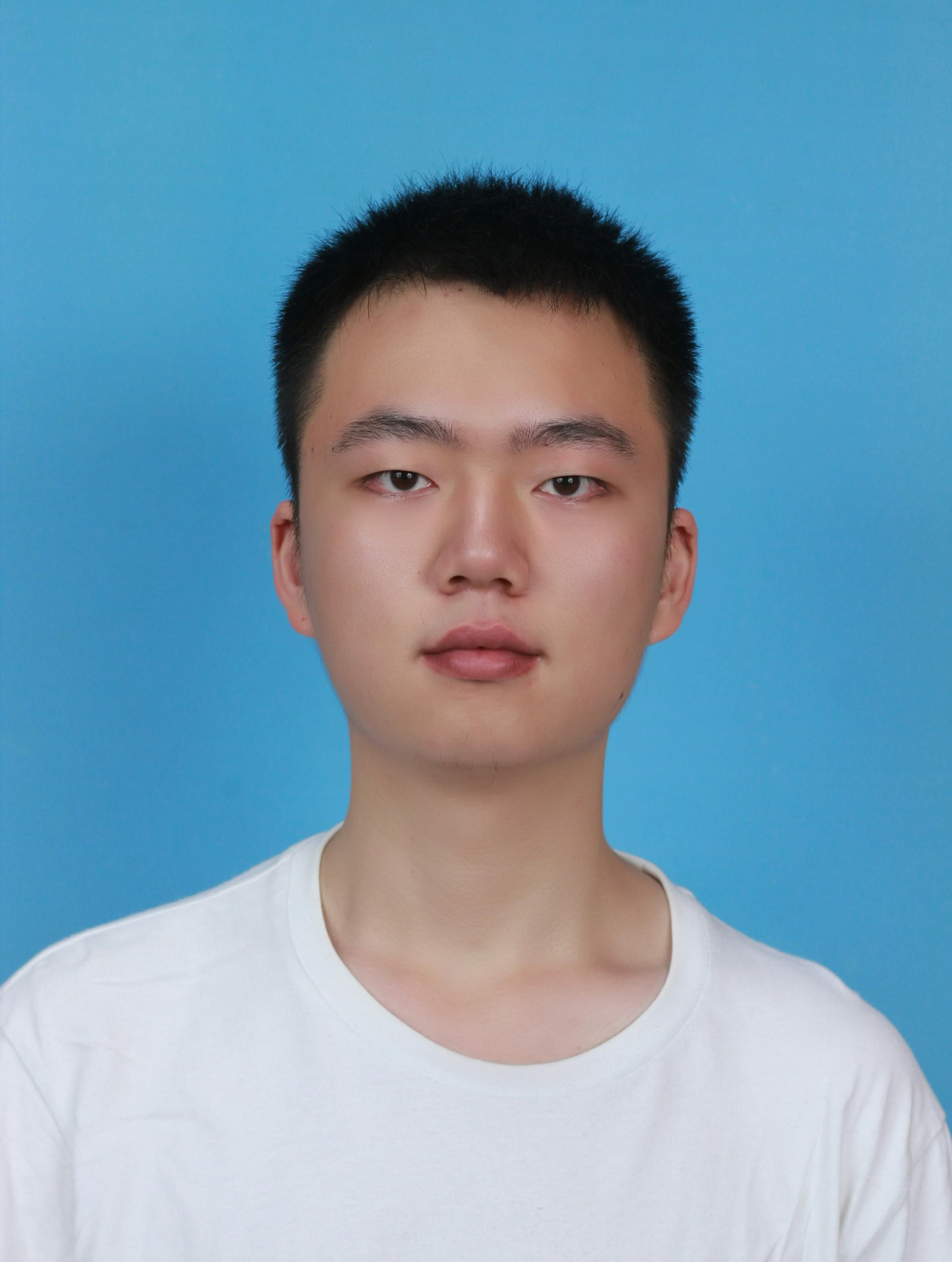}}]{Ye Ke} received the joint bachelor’s degree from the University of Electronic Science and Technology of China and the University of Glasgow. He is currently pursuing the Ph.D. degree with the BRAIN Laboratory, CityU, Hong Kong, working on in-memory computing and brain–machine interfaces.
\end{IEEEbiography}

\begin{IEEEbiography}[{\includegraphics[width=1in,height=1.25in,clip,keepaspectratio]{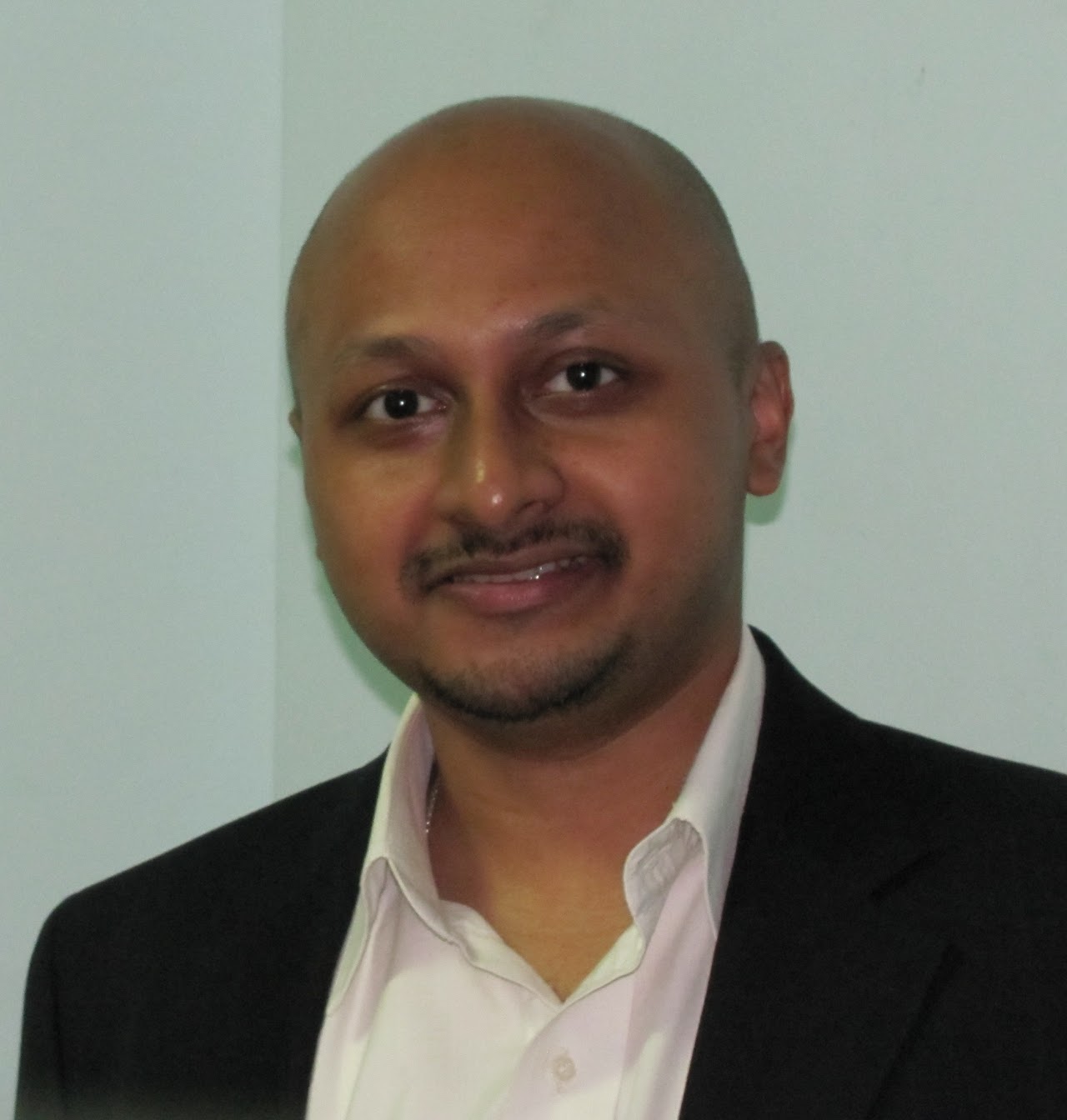}}]{Arindam Basu} (Senior Member, IEEE) received the B.Tech and M.Tech degrees in Electronics and Electrical Communication Engineering from the Indian Institute of Technology, Kharagpur in 2005, the M.S. degree in Mathematics and PhD. degree in Electrical Engineering from the Georgia Institute of Technology, Atlanta in 2009 and 2010 respectively. Dr. Basu received the Prime Minister of India Gold Medal in 2005 from I.I.T Kharagpur. 

He is currently a Professor in City University of Hong Kong in the Department of Electrical Engineering and was a tenured Associate Professor at Nanyang Technological University before this. 

He is currently an Associate Editor-in-Chief of IEEE Transactions on Biomedical Circuits and Systems and an Associate Editor of IEEE Sensors journal, Frontiers in Neuroscience, IOP Neuromorphic Computing and Engineering. He has served as IEEE CAS Distinguished Lecturer for 2016-17 period. Dr. Basu received the best student paper award at Ultrasonics symposium, 2006, best live demonstration at ISCAS 2010 and a finalist position in the best student paper contest at ISCAS 2008. He was awarded MIT Technology Review's TR35 Asia Pacific award in 2012 and inducted into Georgia Tech Alumni Association's 40 under 40 class of 2022. 

He is a technical committee member of the IEEE CAS societies of Biomedical Circuits and Systems, Sensory Systems and Neural Systems and Applications (past Chair). 
\end{IEEEbiography}


\end{document}